\pdfoutput=1

\def\K{$^{39}$K }
\def\Na{$^{23}$Na }

\documentclass[aps,pra,twocolumn,floatfix,10pt]{revtex4} 
\usepackage{graphicx,amsmath,amssymb,ulem}
\usepackage{hyperref}
\usepackage{color}
\usepackage{nicefrac}
\usepackage[caption=false]{subfig}

\begin{document}

\title{Feshbach spectroscopy and dual-species Bose-Einstein condensation of $^{23}\mathrm{Na}-$$^{39}\mathrm{K}$ mixtures}

\author{Torben A.~Schulze}
\author{Torsten~Hartmann}
\author{Kai~K.~Voges}
\author{Matthias~W.~Gempel}
\author{Eberhard~Tiemann}
\author{Alessandro~Zenesini}
\author{Silke~Ospelkaus}

\affiliation{Institut f\"ur Quantenoptik, Leibniz Universit\"at Hannover, 30167~Hannover, Germany}

\date{\today}

\begin{abstract}
We present measurements of interspecies Feshbach resonances and subsequent creation of dual-species Bose-Einstein condensates of $^{23}\mathrm{Na}$ and $^{39}\mathrm{K}$. We prepare both optically trapped ensembles in the spin state $\left|f = 1,m_{f}=-1\right\rangle$ and perform atom loss spectroscopy in a magnetic field range from 0 to $700 \, \mathrm{G}$. The observed features include several s-wave poles and a zero crossing of the interspecies scattering length as well as inelastic two-body contributions in the $\mathcal{M} = m_{\mathrm{Na}}+m_{\mathrm{K}} = -2$ submanifold. We identify and discuss the suitability of different magnetic field regions for the purposes of sympathetic cooling of \K and achieving dual-species degeneracy. Two condensates are created simultaneously by evaporation at a magnetic field of about $150 \, \mathrm{G}$, which provides sizable intra- and interspecies scattering rates needed for fast thermalization. The impact of the differential gravitational sag on the miscibility criterion for the mixture is discussed. Our results serve as a promising starting point for the magnetoassociation into quantum degenerate $^{23}\mathrm{Na}^{39}\mathrm{K}$ Feshbach molecules.

\end{abstract}

\maketitle
\section{Introduction} \label{sec1}
Mixtures of quantum degenerate gases, created by using either different Zeeman sublevels or by using chemically different species, provide a rich testbed for the validation of a plethora of phenomena, ranging from quantum metrology \cite{CAL} to quantum emulation \cite{SOC}. In recent years, the association of ultracold heteronuclear mixtures into diatomic molecules rose to considerable interest. In their absolute ground state, these molecules possess a strong electric dipole moment that enables the study of long-range interacting systems. Amongst others, a characteristic feature expected in strongly dipolar systems is the emergence of supersolidity in an optical lattice environment \cite{Buechler,Lahaye}. Pioneering experiments using the atomic species combination $^{40}\mathrm{K} + ^{87}\text{Rb}$ were the first to demonstrate the nowadays considered default strategy for producing ground state molecules, starting with two quantum degenerate atomic clouds and performing ground state STIRAP after they have been magnetoassociated into Feshbach molecules \cite{KRb1,KRb2,Ospelkaus2010}. A multitude of different alkali combinations are investigated experimentally using this outlined path, including LiNa \cite{LiNa}, RbCs \cite{RbCs}, NaRb \cite{NaRbREVISION} and LiCs \cite{LiCs}, as well as more exotic combinations involving alkaline earth as well as rare earth elements \cite{RbYb}.

Along the lines of the available alkalis, the chemically stable NaK with its electric dipole moment of 2.72 Debye is a promising candidate for its use in such a molecular experiment. Due to its long history of spectroscopic studies \cite{Barratt1924,Barrow,Kato,Zemke,Stolyarov,Russier,Gerdes2008}, ground and excited state potential energy curves of NaK have been accurately determined. Using the natural isotopes $^{39,40,41}\mathrm{K}$, the different NaK combinations provide the unique possibility to switch between the study of fermionic and bosonic systems. In seminal experiments at the MIT, dual quantum degenerate samples using $^{23}\mathrm{Na}$ and the fermionic $^{40} \mathrm{K}$ have been created, associated into molecules and prepared in their absolute ground states \cite{Zwierlein1,Zwierlein2,Zwierlein3}. Recently, Feshbach resonances and STIRAP have been thoroughly studied theoretically and experimentally also by other groups \cite{Bo,Frauke}. Both bosonic combinations, $^{23}\mathrm{Na}^{39}\mathrm{K}$ and $^{23}\mathrm{Na}^{41}\mathrm{K}$, have been left unexplored in the ultracold regime up to now. In this article, we present our measurements of Feshbach resonances in the Bose-Bose mixture as well as the first experimental demonstration of a quantum degenerate mixture of \Na and $^{39}\mathrm{K}$. To its end, we investigate the dual-species collisional properties both theoretically and experimentally, and discuss them in combination with their respective homonuclear properties. \\
We start our investigation with the theoretical framework, which is presented in section \ref{sec:theo}. Our calculations, using recent data of the isotope pair $^{23}\mathrm{Na}^{40}\mathrm{K}$, outline the expected signatures and provide landmarks for the following Feshbach spectroscopy. After a brief description of our experimental setup in section \ref{sec:experiment}, we display our measurements of strong three-body recombination losses at $B = 0\,\mathrm{G}$ in dual-species operation, in agreement with the large, negative value of the scattering length predicted in the theoretical part of this work. In section \ref{sec:Fesh}, we show the measured Feshbach spectrum for the incident channel of interest, in which the exact positions and widths of the calculated signatures are measured. In section \ref{sec:dual}, we combine the heteronuclear with the homonuclear Feshbach resonance spectrum, giving rise to interaction domains, which we discuss with respect to the mean-field stability of a dual-species condensate. Having identified a suitable domain, we show that forced evaporation in our dipole trap leads to a dual-species Bose-Einstein condensate revealing bimodal distributions in time of flight images of both species.

\subsection{Theoretical model and expected signatures}\label{sec:theo}
\begin{figure}[t]
\centering
\includegraphics*[width=\columnwidth]{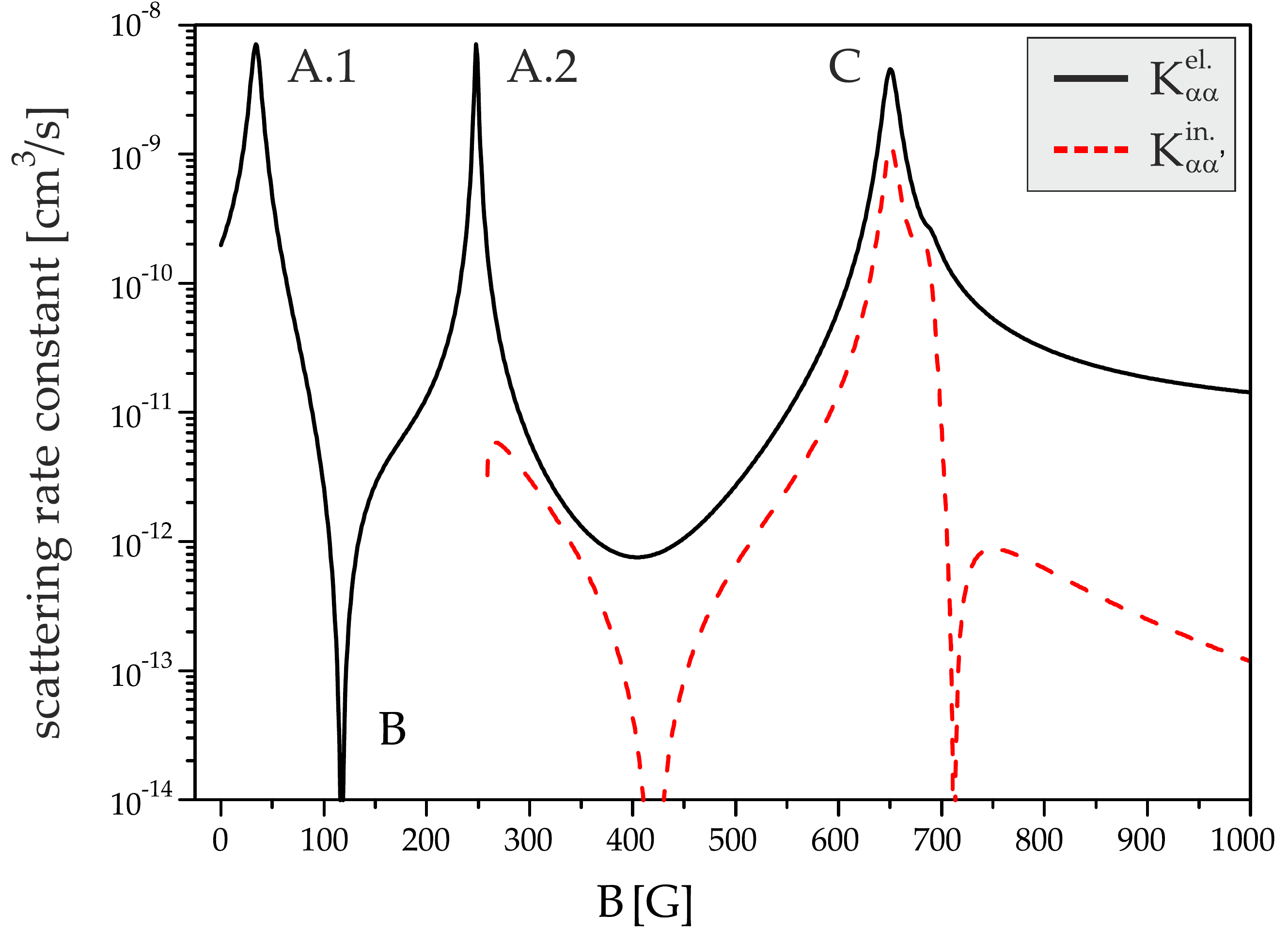}
\caption{Elastic (straight black) and inelastic (dashed red) two-body scattering rate constants as a function of the magnetic field for the incident channel $\alpha$. The labels A, B and C denote Feshbach resonances, interaction zeroes and channel mixing, respectively.}
\label{fig0}
\end{figure}

The Hamiltonian $H$ for the diatomic collision reads
\begin{equation} H = T + V + H_{\text{hf}} + H_{\text{dd}} + H_{\text{Zee}} \label{eq:tiemann} \end{equation}
with T being the kinetic energy operator and $H_{\text{hf,(dd),[Zee]}}$ the hyperfine (magnetic spin-spin) [Zeeman] interaction, respectively. V is given by the singlet ($S = 0$) and triplet ($S = 1$) potential energy curves (PECs) and corresponding projectors, $V = P_{0}V_{0} + P_{1}V_{1}$. We use the PECs obtained in \cite{Gerdes2008} and refine them using data from a molecular beam study on thermal $^{23}\mathrm{Na}^{39}\mathrm{K}$ \cite{Temelkov} and data from Feshbach spectroscopy on $^{23}\mathrm{Na}^{40}\mathrm{K}$ presented in \cite{Zwierlein1} and most recently in \cite{Bo}. The resonances measured in \cite{Zwierlein1} have mainly triplet character and will therefore primarily refine the $a^{3}\Sigma^{+}$ potential, whereas larger deviations might persist for the singlet part. The new evaluation in \cite{Bo} gave a significant change for the singlet potential, because their observed resonances (s and d) involve bound states with significant singlet character. Using the refined PECs, we then perform isotopic rescaling by changing the reduced mass $\mu$ in $T$ of Eq.\,\eqref{eq:tiemann} as well as appropriate hyperfine parameters and g-factors for $^{39}\mathrm{K}$. By numerically solving the corresponding coupled-channel Schroedinger equation using a Numerov algorithm at a given collisional energy, we obtain the resonance-dependent collisional properties such as scattering lengths and rates out of the scattering phase shift. The scattering channels are written as $(\left|f,m_{f}\right\rangle_{\mathrm{Na}} \left|f,m_{f}\right\rangle_{\mathrm{K}} \left|\ell,m_{\ell}\right\rangle)$, where $f$ is given by the total angular momentum of the respective atom, $\ell$ by their relative angular momentum and $m_{f},m_{\ell}$ denote their projection onto the quantization axis. In our experiment, we are preparing the pair $|1,-1\rangle_{\mathrm{Na}} + |1,-1\rangle_{\mathrm{K}}$. We define $\alpha = \left|1,-1 \right\rangle_{\text{Na}} \left|1, -1 \right\rangle_{\text{K}}\left|0,0\right\rangle$ as our entrance channel, which is dominant because of the low temperature. Similarly, we restrict our Hilbert space to $\ell = 0$, and therefore do not include the appearance of d-wave resonances or inelastic collisions due to dipole-dipole interaction. Figure\,\ref{fig0} shows the elastic and inelastic two-body scattering rate constants $K^{\text{el.}}_{\alpha \alpha}(B)$ and $K^{\text{in.}}_{\alpha \alpha'}(B)$
for the collision space $\mathcal{M} = m_{\mathrm{Na}} + m_{\mathrm{K}} + m_{\ell} = -2$ in a magnetic field region from 0 to 1000 G and a collisional energy of $1 \, \mathrm{\mu K}$. At least three different signatures can be identified, which we denote as A, B and C for convenience: For the incoming spin state combination, Feshbach resonances (A.1 \& A.2) are expected at $34.2 \, $G and $248.1 \, $G, as shown by the local maxima of the elastic two-body scattering. Similarly, a zero crossing (B) of the scattering length exists at $117.3 \, $G. Finally, at about $259 \,$G, another channel of the $\mathcal{M} = -2$ space, namely $m_{\text{Na}}=0$ and $m_{\text{K}}=-2$, opens up by crossing the threshold $E_{|\alpha\rangle}$. At magnetic fields larger than 259\,G, transitions to the now open channel give rise to the inelastic part of the incident channel, displaying a peak (C) at $651 \,$G, with a shoulder at $690\,$G. We refer to section \ref{sec:Fesh} for a more detailed discussion of the last feature. As all three of these signatures will evoke or impede losses in the mixture, all of them can be located using atom loss spectroscopy. We note that the low-field resonance differs by about $30\,$G from the one predicted in \cite{VielSimoni}. If we do not include the refinement procedure from the work of \cite{Bo}, the calculated low-field resonance shifts significantly to lower fields. Measuring the exact location of this resonance will therefore be imperative, as it gives direct insight into the singlet component. It further dictates the exact value of the interspecies scattering length $a_{\text{NaK}}$ in absence of magnetic fields. From the calculations shown in Fig.\,\ref{fig0}, $a_{\text{NaK},|\alpha\rangle}(B = 0) = -416 \; a_{0}$ for $|\alpha\rangle =  \left|1, -1 \right\rangle_{\text{Na}} \left|1, -1 \right\rangle_{\text{K}}\left|0,0\right\rangle$, where $a_{0}$ is the Bohr radius. This large value will aid sympathetic cooling mechanisms in the low density stage of the experiment, but also produce losses in the high phase-space density environment of an optical dipole trap. We return to this in section \ref{sec:B0}.

\subsection{Experimental setup}\label{sec:experiment}
%
%
Our experimental apparatus is based on the one described in \cite{Gempel}. It combines two pre-cooled atomic sources into an UHV collection region, where a two-color magneto-optical trap (MOT) is operated. \Na is heated up in an oven region and slowed down using a spin-flip Zeeman slower. The atom number is extracted by recording the MOT fluorescence, giving $4 \times 10^9$ atoms in the 3D-MOT. \K is distributed using commercial dispensers, and pre-cooled using a two-dimensional MOT with an additional pushing beam along the longitudinal axis. Bosonic \K as well as fermionic $^{40}$K MOTs have been realized in this setup. For the bosonic case, the MOT contains $1 \times 10^{8}$ \K atoms in single-species and $6.4 \times 10^{7}$ atoms in dual-species operation. \Na and \K are further sub-Doppler cooled for 5 ms \cite{Landini} and optically pumped into their respective low-field seeking state $\left|f = 1, m_{f} = -1 \right\rangle$. Afterwards they are captured in a magnetic quadrupole trap with a magnetic field gradient of 216 G/cm along the strongly-confining vertical direction. A blue-detuned laser beam ($\lambda_{\mathrm{BP}} = 532\,\mathrm{nm}, P = 3.25\,\mathrm{W}, w_{0} = 39 \, \mathrm{\mu m}$) propagating along the symmetry axis of the trap gives rise to a repulsive potential at the magnetic field zero in the trap center, serving as an optical plug of the magnetic trap bottom. In single-species operation, the lifetime of the individual samples in the plugged trap exceeds $45\,$s, allowing for evaporative cooling using either radio frequency or microwave fields. We evaporate solely the \Na ensemble by selectively removing the hottest atoms from the trap through microwave transitions to the high-field seeking $f = 2$ states. In dual-species operation, a small fraction of \K is added and sympathetically cooled via $^{23}\mathrm{Na}$. At the end of the 12\,s long evaporation stage, the phase space compressed sample is transferred into a $90^{\circ}$-crossed optical dipole trap (cODT) consisting of two far-detuned ($\lambda_{\mathrm{ODT}} = 1064 \, \mathrm{nm}$) overlapping laser beams in the horizontal x-y plane. The optical trap combines a tightly confining (4.72 W, $w_{0} = 46 \, \mathrm{\mu m}$) beam ensuring high optical trap depth and an elliptically shaped (3.83 W, $(w_{0,\mathrm{y}},w_{0,\mathrm{z}})=(143\, \mathrm{\mu m} \times 40\, \mathrm{\mu m})$) beam enlarging the trap volume. The temperature of the two atomic clouds in the cODT is $T = 2.3 \, \mu \mathrm{K}$, wheras the individual atom numbers are tuned in a range of $10^{4}$ to $10^{6}$ depending on the atom number imbalance that we want to prepare. A residual guiding field $B_{\mathrm{res}} = 2.49 \, \mathrm{G}$, applied along the vertical direction, preserves the individual spin projection of the optically trapped atoms, $m_{f,\mathrm{Na}} = m_{f,\mathrm{K}} = -1$.
\subsection{Collisional properties at $B = 0 \, \mathrm{G}$.}\label{sec:B0}
\begin{figure}[t]
\includegraphics[width=\columnwidth]{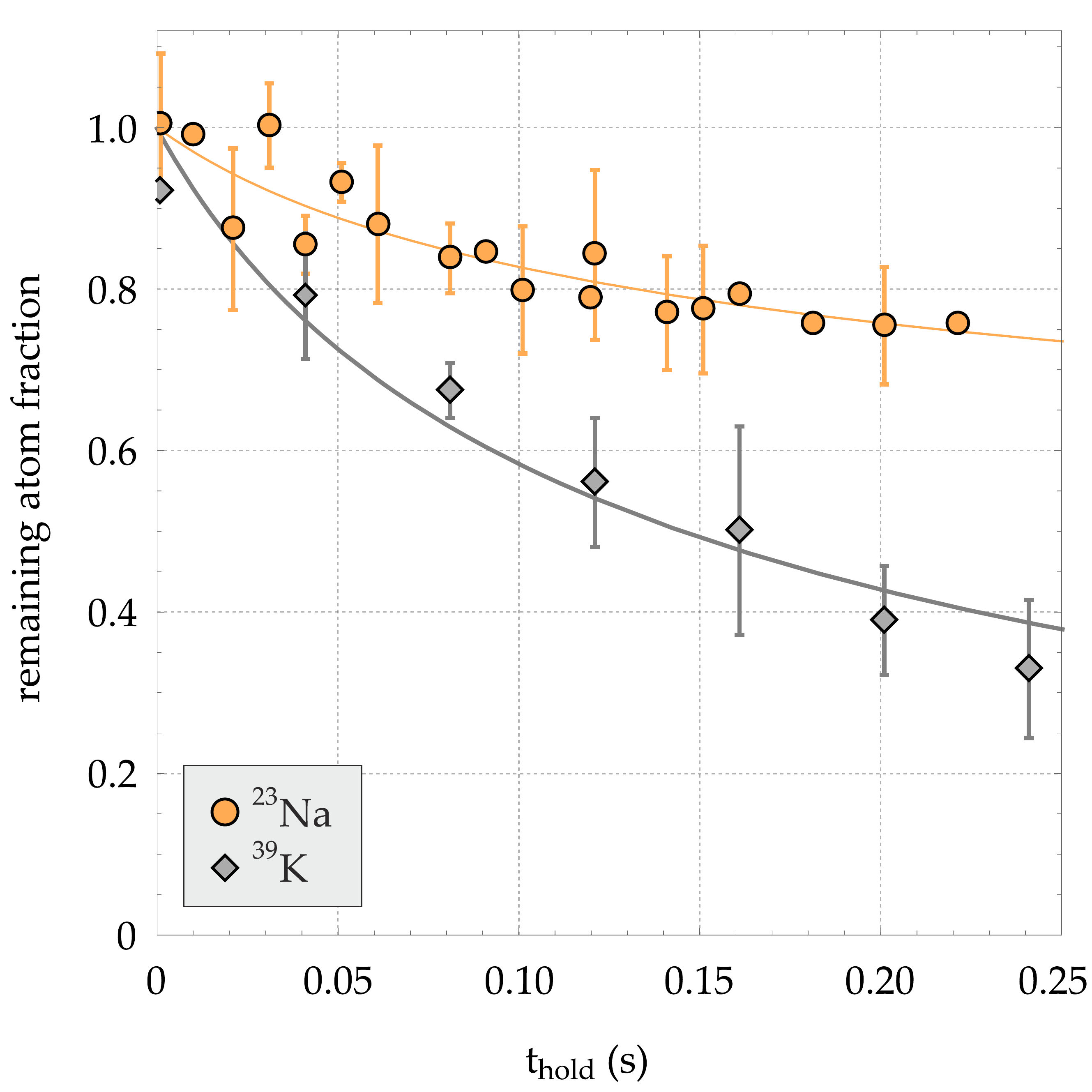}
\caption{Remaining fraction of \Na (orange circles) and \K (gray diamonds) atoms as a function of the holding time in the dipole trap at $B = 0 \, \mathrm{G}$. Three-body losses lead to a rapid atom number decay of the minority component $^{39}\mathrm{K}$. The solid line is a fit of the coupled-system loss rate equations solution to the acquired data. Both atom numbers are normalized to their fitted value at t = 0, being $3.8\times 10^{5}$ for \Na and $1.1 \times 10^{5}$ for $^{39}\mathrm{K}$.}
\label{fig1}
\end{figure}

Due to the high densities prevailing in the cODT, the atomic clouds can experience sizable three-body recombination effects, with the three-body loss coefficient scaling with the fourth power of the appropriate scattering length \cite{efi}. To quantify the different loss contributions, we first extract the single-species collisional loss properties of $^{39}\mathrm{K}$. Following a removal of the \Na cloud via a resonant light pulse before the cODT transfer, we monitor the potassium atom number as a function of the holding time $t_{\mathrm{hold}}$ in the dipole trap. We do not observe a considerable amount of three-body losses, and extract a 1/e lifetime of the sample of 37.8 seconds which mainly originates from collisions with background gas atoms. We then repeat the experiment with the mixture at a temperature of $2.6\,\mu \mathrm{K}$ and an imbalanced atom number ratio $N_{\mathrm{Na}}/N_{\mathrm{K}} \gg 1$. The atom number as a function of the holding time for both species is depicted in Fig.\,\ref{fig1}.
	Rapid losses are observed in the \K signal, with an effective 1/e lifetime of about $240 \, \mathrm{ms}$, whereas $^{23}\mathrm{Na}$, being the majority cloud, shows only a slight decrease over time. We extract the three-body loss coefficients $L_{\mathrm{Na},\mathrm{Na},\mathrm{K}}$ and $L_{\mathrm{Na},\mathrm{K},\mathrm{K}}$ by fitting the experimental data to the solutions of three coupled differential equations governing the sodium and potassium atom number loss dynamics as well as the temperature time evolution \cite{Wacker1, efi}; see Appendix. The temperature increases because losses occur more frequently in the high-density region of the trap center (anti-evaporation). We do not account for competing evaporation effects in the samples, because the temperature increase of about 30\,\% is well below the trap depth. We also neglect recombination heating as it becomes critical only at positive scattering length, where the probability that the recombination product remains trapped considerably increases. The obtained loss coefficent are $L_{\mathrm{Na},\mathrm{Na},\mathrm{K}}=1.03(62)\times 10^{-25}\, \, \mathrm{cm}^{6}\mathrm{s}^{-1}$ and $L_{\mathrm{Na},\mathrm{K},\mathrm{K}}=0.50(30)\times 10^{-25}\, \mathrm{cm}^{6}\mathrm{s}^{-1}$, where the uncertainties include the statistical error on the fit as well as systematic uncertainties in the evaluation of the experimental quantities: temperature, atom number and trap frequencies for both species. The relation between three-body recombination rates and two-body scattering length depends on non-trivial Efimov physics \cite{Efimov1970ela, Kraemer2006efe} and consequently on the presence or absence of resonant structures for the studied mixture. Yet comparison with extensive analysis on other alkali mixture indicates that the loss rate coefficients are on the order of the ones observed in $^{39}\mathrm{K}^{87}\mathrm{Rb}$ and $^{41}\mathrm{K}^{87}\mathrm{Rb}$ \cite{Wacker1}, $^{40}\mathrm{K}^{87}\mathrm{Rb}$ \cite{bloom2013} and $^{6}\mathrm{Li}^{133}\mathrm{Cs}$ \cite{Pires2014, Tung2014} for scattering lengths of few hundred $a_0$. Our extracted loss coefficients therefore appear to be compatible with our theoretically calculated value of $a(B=0) = -416 \, a_{0}$. Future investigations on three-body loss rates at different scattering lengths will allow a better understanding of the underlying scenario in this specific mixture. With the high amount of undesirable losses, $B = 0 \,$G does not provide a suitable environment for operating the dual-species apparatus beyond a low-density region. In order to impede three-body losses, the interspecies scattering has to be reduced, for example by using magnetic fields.

\section{Feshbach spectrum}\label{sec:Fesh}
\subsection{Methods}
We perform heteronuclear Feshbach spectroscopy by preparing an atom-number imbalanced mixture $(N_{\text{major}}/N_{\text{minor}} \approx 10)$ through adjustment of the individual MOT loading times, and searching for spectral features as a function of the strength of an applied magnetic bias field $B$, using the minority cloud as a probe while the majority component serves as a bath. For the spin state under consideration, $\left|f = 1, m_{f} = -1\right\rangle$, both \Na and \K have been investigated and Feshbach resonances located. For $^{23}\mathrm{Na}$, the lowest magnetic field resonance in this particular spin state appears at $1195 \, \mathrm{G}$ \cite{Ketterle}, far beyond the magnetic field region $B \leq 700 \, \mathrm{G}$ probed in this experiment. Homonuclear \K collisions exhibit three s-wave resonances, which are situated at $32.6\, \mathrm{G}$, $162.8\, \mathrm{G}$ and $562.2\, \mathrm{G}$, respectively \cite{K39feshbach}. Pure \K Bose-Einstein condensates have been realized in the vicinity of all these resonances \cite{AspectK39}. In order to distinguish between homonuclear and heteronuclear effects, we record the pure \K resonance spectrum in addition to the dual-species loss spectrum. 
Our magnetic fields are calibrated using microwave spectroscopy of the \Na cloud at $700-800\, \mathrm{nK}$. For a given electric current producing the magnetic field, we measure microwave transitions $\left| f = 1, m_{f} = -1 \right\rangle \rightarrow \left| f = 2, m_{f} = 0 \right\rangle$ and calculate the corresponding magnetic field using the Breit-Rabi-formula. The microwave dip positions are determined with a statistical uncertainty on the order of $10 \, \mathrm{kHz}$. For a given bias field of $100 \, \mathrm{G}$, this translates into a magnetic field uncertainty of $34\, \mathrm{mG}$. In our experimental sequence, the bias field is first ramped up in 5 ms to a magnetic field value of $B_{0}\approx 100 \, \mathrm{G}$, where it is held for less than half a second in order to ensure that the samples are thermalized. This magnetic field window serves as a safe spot for the dual-species operation, as neither the inter- nor the intraspecies interactions are disruptive in this region (see also Fig.\,\ref{fig5}). A rigorous discussion of this region follows in section \ref{sec:dual}. 

Starting from $B_{0}$, the bias field is ramped in a few milliseconds to a specific value $B_{f}$ and held there for variable holding times $t_{\text{hold}} \geq 100\,\mathrm{ms}$. At the end of each cycle, all magnetic fields are switched off and the clouds are released from the cODT. Atom numbers are obtained by performing absorption imaging, where the majority is measured \textit{in-situ} and the minority in time of flight. As soon as a spectral variation in the atom number is found, the holding time is adjusted in order to obtain a clearly visible drop in the respective signals without depleting one of the atomic samples completely. A magnetic field scan using this fixed holding time then recovers the corresponding spectral feature.
The expected heteronuclear signatures are the ones outlined in Fig.\,\ref{fig0}. 
\subsection{Experimental results}
\begin{figure}[bht]
\centering
\includegraphics[width=\columnwidth]{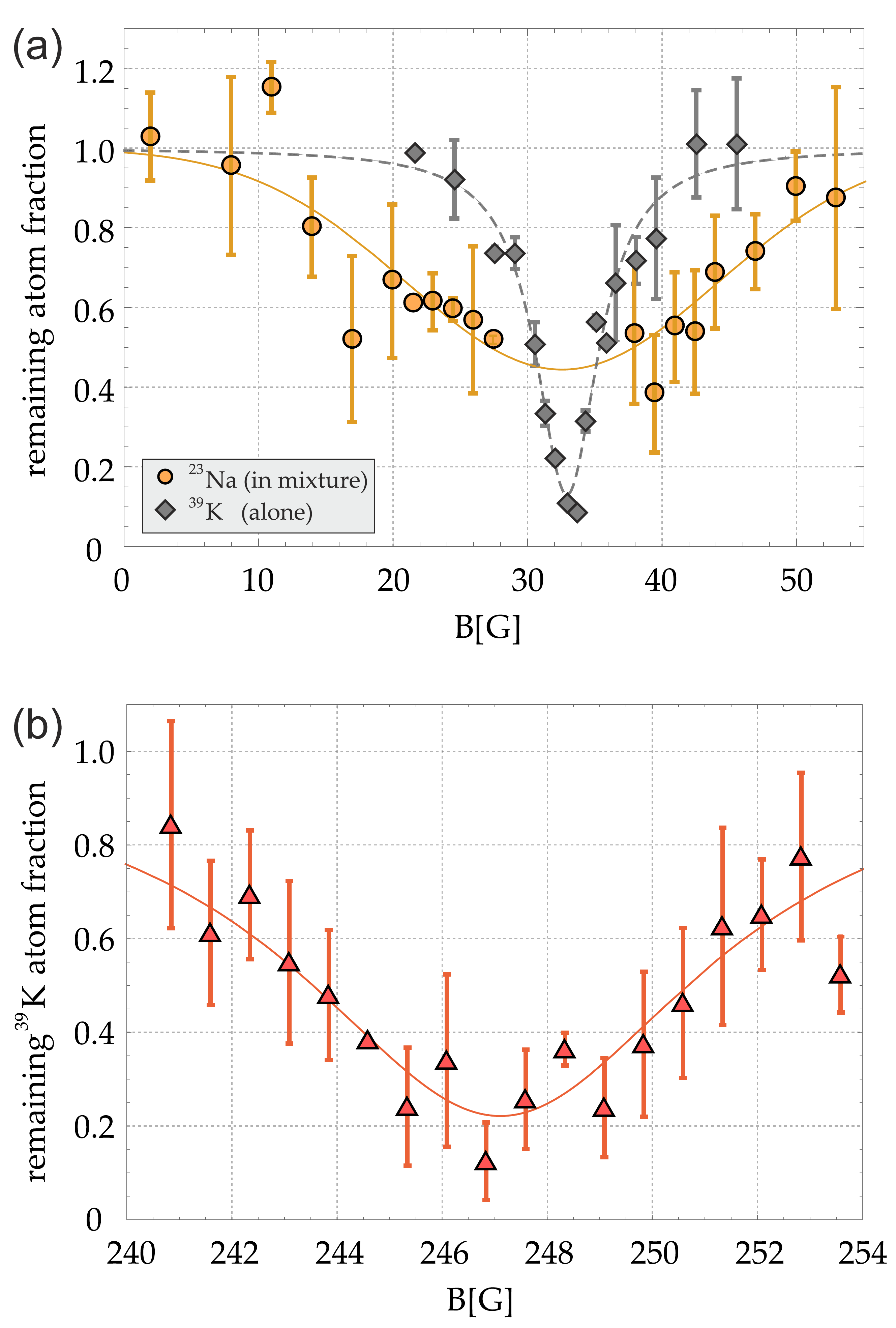}
\caption{Remaining fraction of atoms as a function of the applied magnetic field strength $B$. The Feshbach resonances A.1 and A.2. are visible due to drops in the atom number. (a) \K Atom number loss in single-species operation (gray diamonds) and \Na losses in dual-species operation (yellow dots) unveil two overlayed resonances. The fitted function (dashed line) determines the homonuclear resonance. A Gaussian fit (solid line) to the wings of the \Na minority signal determines the heteronuclear resonance location. The absolute atom numbers are $3 \times 10^4$ ($^{23}\mathrm{Na}$) and $1.5 \times 10^5$ ($^{39}\mathrm{K}$). (b) The high-field resonance A.2, measured using \K as the minority. The absolute atom number is $2.6 \times 10^5$.}
\label{fig:2a} 
\end{figure}
Figure\,\ref{fig:2a} shows the recorded atom losses corresponding to signature A.1 (a) and A.2 (b). The low-field resonance A.1 around $30\, \mathrm{G}$ was found to be strongly interwoven with a homonuclear $^{39}\mathrm{K}$ resonance. To discern the overlapping resonances, we make use of the fact that they exhibit largely different widths. The homonuclear resonance is spectrally sharp with a comparatively small width of $\sigma_{\mathrm{K}} = 5.7 \, \mathrm{G}$, where $\sigma$ denotes the full width half maximum value of the loss signal. For the heteronuclear collision, the large negative background scattering length is equivalent to the existence of a virtual state right above threshold. Large continuum coupling persists in this channel, which will in turn broaden the resonance strongly ($\sigma_{\mathrm{A.1}} = 27.3 \, \mathrm{G}$). We first measure the homonuclear resonance (gray diamonds in Fig.\,\ref{fig:2a} (a)). Our fit gives a location of $32.9\, \mathrm{G}$, being well in the error bars of \cite{K39feshbach}, where $32.6 \, \mathrm{G}\pm 1.5 \, \mathrm{G}$ was measured. The heteronuclear signature is recorded using \Na as the minority probe. We determine a region spanning $\approx2\sigma_{\mathrm{K}}$ in which the heteronuclear signature is affected by the \K resonance, and filter it out for our fitting procedure of the A.1 resonance location. Even without the central part of the feature\cite{Amplitude}, we can locate it with a statistical uncertainty of $0.8\, \mathrm{G}$ to be at $32.5\,\mathrm{G}$. No complication exists for the high-field resonance A.2, which is measured to be at $247.1 \, \mathrm{G} \pm 0.2 \, \mathrm{G}$ using \K as the minority probe. Both locations are in agreement with our theoretical predictions (see Fig.\,\ref{fig0}), with small deviations on the order of one Gauss.

\begin{figure}[hbt]
\centering
   \includegraphics[width=\columnwidth]{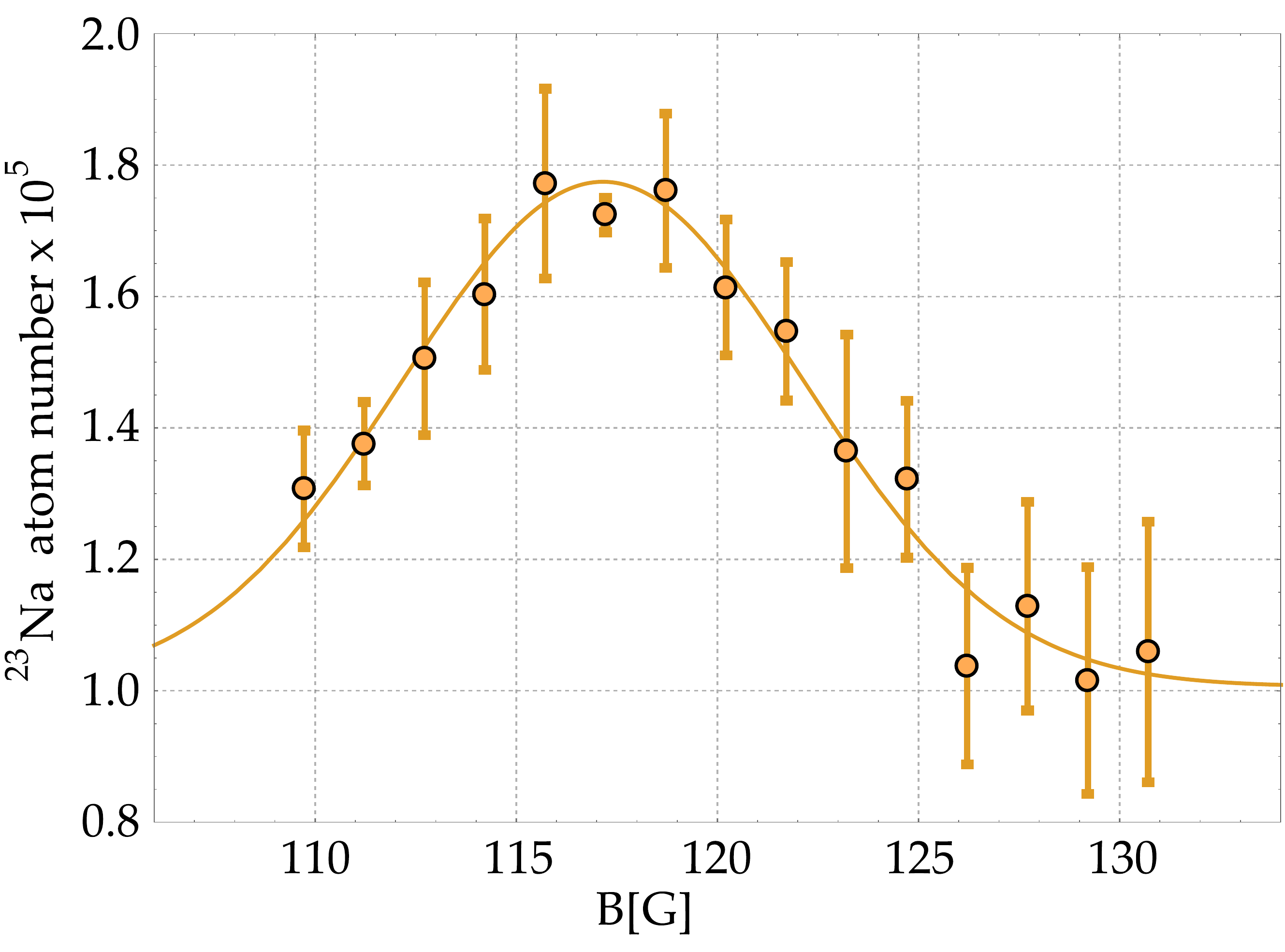}
\caption{\Na atom number as a function of the magnetic field strength $B$ in a forced evaporation sequence. The presence of \K leads to \Na loss enhancement via sympathetic cooling. A revival of the \Na signal at $\sim 117 \, $G indicates the reduction of this loss enhancement as the interspecies scattering rate goes to zero. A Gaussian fit (solid line) is used to extract its location.} \label{fig:TZC}
\end{figure}

We now turn to the remaining signatures. Figure\,\ref{fig:TZC} shows the part of the acquired spectrum in which the zero point crossing (signature B) was investigated. The value at which the interspecies scattering turns from repulsive to attractive is of particular interest for the dual-species operation, as it provides the magnetic field region in which atom losses are expected to be low. Note that in general the position of the two-body scattering zero crossing will not be identical to the minimum of three-body losses, and a measurement similar to sec. \ref{sec:B0} can give misleading results. We instead localize this field position by exploiting the two-body losses that appear during optical evaporation. For the optical trapping potential $U_{\mathrm{i}}$, where i denotes the species, one finds $U_{\text{K}} \approx 2.51$ $U_{\text{Na}}$, i.e. a preferable ejection of \Na as the cODT intensities are reduced. Sympathetic cooling of \K by \Na will lead to an enhancement of \Na losses during rethermalization. In our experiment, we first prepare an atom number balanced mixture at $\sim 10 \, \mathrm{\mu K}$. After lowering the optical potential depth in $1.5 \, \mathrm{s}$ to a value which gives $2 \, \mathrm{\mu K}$ in single-species \Na operation, we record the sodium atom number after a thermalization time. The loss enhancement will be reduced and ultimately disabled as the interspecies scattering rate approaches zero, leading to a sodium signal revival. By fitting a phenomenological Gaussian to the observed signal in Fig.\,\ref{fig:TZC}, we determine the zero-crossing to be located at $117.2 \, \mathrm{G} \pm 0.2 \, \mathrm{G}$, in agreement with our predicted value of 117 G (see Table\,\ref{tab:table1}). 
\begin{figure}[hbt]
   \includegraphics[width=\columnwidth]{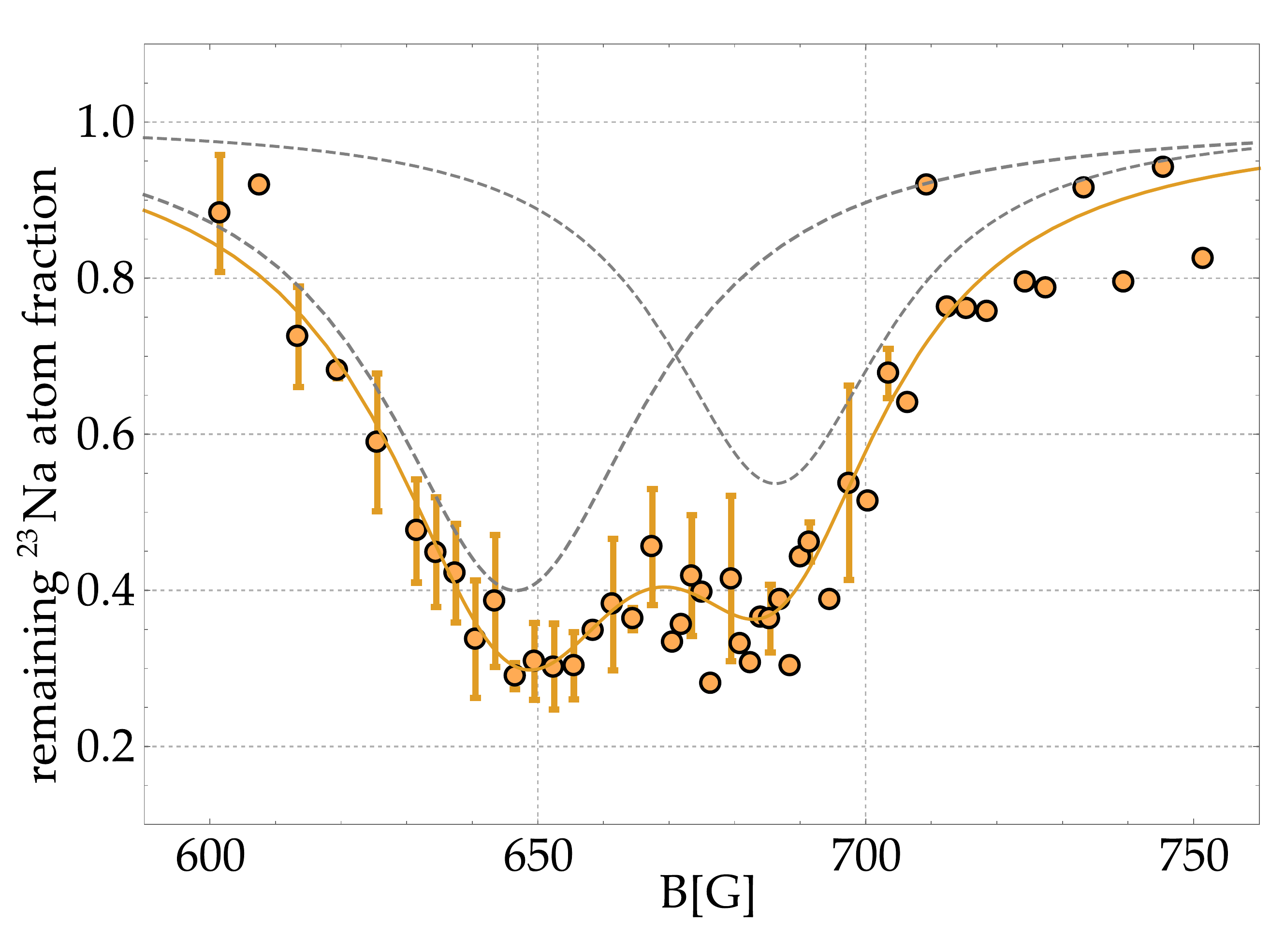}
   \label{fig:Ng2}
\caption{Remaining fraction of \Na atoms as a function of the applied magnetic field strength $B$. Two overlapping loss features are discerned, which are identified as signature (C) from Fig.\,\ref{fig0}. A combined fit (straight line) is given together with two individual Lorentzian loss profiles (dashed lines). The absolute atom number is $3.7 \times 10^{5}$ \Na atoms.}
\label{fig:doubl}
\end{figure}

The loss feature in Fig.\,\ref{fig:doubl}, previously denoted as C, is the result of channel mixing between the states spanning the manifold of $\mathcal{M} = -2$. At a magnetic field of $259 \,\mathrm{G}$, the state $|\beta\rangle = \left|f = 1, m_{f}=0 \right\rangle_{\text{Na}} + \left|f = 2, m_{f}=-2 \right\rangle_{\text{K}} $ becomes an open channel by crossing the collisional threshold of the incident channel $|\alpha\rangle$. Starting from that field, $|\alpha\rangle$ will be embedded in the continuum of $|\beta\rangle$, and inelastic collisions $|\alpha\rangle \rightarrow |\beta\rangle$ are energetically allowed. This effect becomes sizable in the magnetic field region between 650 and 700 G, when two molecular states cross $|\alpha\rangle$ and resonantly enhance the scattering rate constant. Any collisional decay to $|\beta\rangle$ is accompanied by a kinetic energy gain $E_{\alpha \beta} = E_{\alpha}(B)-E_{\beta}(B)$ of the atom pair. In the region $B > 600 \, \mathrm{G}$, $E_{\alpha \beta} > 9.7 \, \mathrm{mK}$ strongly exceeds the optical confinement, and the atom pair leaves the trap. The channel mixing process can therefore be directly probed using atom loss spectroscopy. As losses in this resonance region will be primarily mediated by two-body instead of three-body collisions, dynamics on faster timescales are expected. We hold the sample at a given magnetic field for $t_{\text{hold}} \leq 10 \, \mathrm{ms}$, much shorter than the holding times for signatures A.1 and A.2. We observe almost complete depletion of the atomic signal, with a $1/\mathrm{e}$ lifetime of $4\, \mathrm{ms}$ at $650 \, \mathrm{G}$. In unison with the calculations presented in section \ref{sec:theo}, the strongly shortened effective lifetime suggests two-body collisions as the driving loss mechanism. This assumption can be further tested by examining the atom number decay at different densities or temperatures \cite{2body}, which is not investigated here. The observed double feature corresponds to the predicted overlap of two Feshbach resonances (see shoulder in the inelastic contribution in Fig.\,\ref{fig0}). We extract their individual positions by a joint fit as shown in Fig.\,\ref{fig:doubl}.
In total, we have located the s-wave resonances A.1 and A.2, the zero crossing B and the two closely spaced resonances C. Their positions and widths, together with our calculations and recent predictions of \cite{VielSimoni} are summarized in Table\,\ref{tab:table1}. 
\begin{table}[bth]
\begin{ruledtabular}
\begin{tabular}{l l l l l}
Feature & $B_{\text{Exp.}}$ & $\sigma_{\text{Exp.}}$ & $B_{\text{Th.}}$\footnote{\text{Theoretical calculations using the data presented in \cite{Bo}.}} & $B_{\text{Th.}}\text{\cite{VielSimoni}}$\\
A.1 & 32.5 (0.8) & 27.3 (2.9) & 34.2 & 2.0\\
A.2 & 247.1 (0.2) & 9.5 (3.0) & 248.1 & 241.4\\
B & 117.2 (0.2) & 11.7 (1.1) & 117. & 75.7 \\
C & 646.6 (1.5) & 48.6 (5.6) & 651.5 & -\\
& 686.2 (1.5) & 40.9 (5.9) & 686.7 & -\\
\end{tabular}
\caption{Experimental magnetic field positions $B_{\text{Exp.}}$, FWHM widths $\sigma_{\text{Exp.}}$ and respective statistical standard errors $(\pm)$ together with the theoretically calculated positions $B_{\text{Th.}}$ of this article and a comparison with recent work \cite{VielSimoni}.\label{tab:table1}}
\end{ruledtabular}
\end{table}
\begin{figure*}[tb]
\centering
\includegraphics*[width=2\columnwidth]{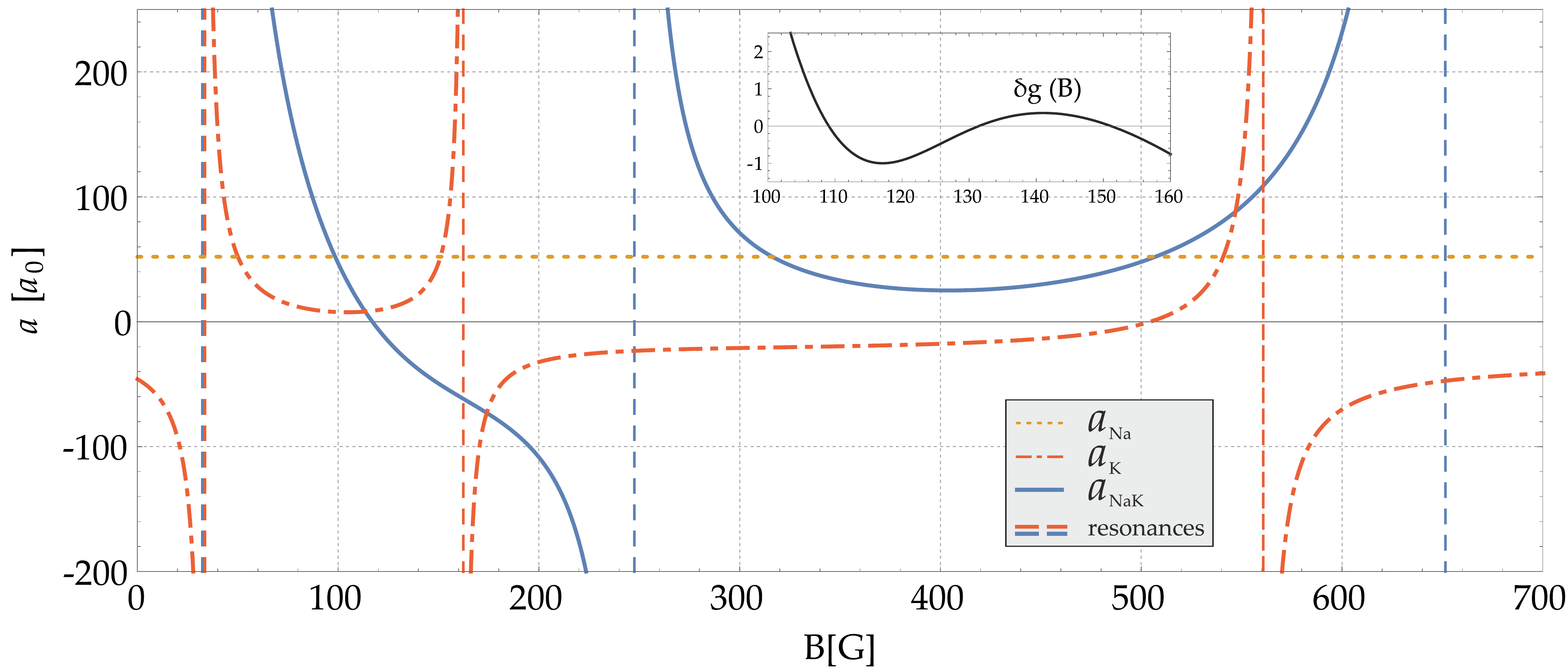}
\caption{Inter- and intraspecies scattering lengths $a_{\text{NaK}}$, $a_{\text{Na}}$ and $a_{\mathrm{K}}$ as a function of magnetic field strength. The positions of \K and $^{23}\mathrm{Na}^{39}\mathrm{K}$ resonances are indicated by dashed vertical lines. Inset: Miscibility parameter $\delta g$ as a function of magnetic field. Three sign changes exist in the magnetic field valley outlined by the two low-field $^{39}\mathrm{K}$ resonances.}  
\label{fig5}
\end{figure*}

\section{Tunability of Interactions}\label{sec:dual}
\subsection{Interaction domains}
Based on the recorded Feshbach spectrum, we are now able to discuss magnetic field regions suitable for producing quantum degenerate atomic samples of two species. To this end, we search for a domain in which either both species can efficiently cool themselves and display a non destructive interspecies behaviour, or in which one of the samples can efficiently cool the other one sympathetically. Figure\,\ref{fig5} shows the interspecies scattering length from 0 to $700 \, \mathrm{G}$ based on the results obtained in the last section, together with the intraspecies scattering taken from \cite{NaScattering} for \Na and \cite{K39feshbach,simoniprivate} for $^{39}\mathrm{K}$. Due to the magnetic field independent interaction of $^{23}\mathrm{Na}$ in this magnetic field window, it can be condensed at arbitrary bias field values, and its scattering length $a_{\text{Na}} = 52\,a_{0}$ will only contribute to a positive interaction offset on the dual-species operation. For \K on the other hand, the appearance and relative shape of its Feshbach resonances constrains our magnetic field of operation to a magnetic field valley between $32.6 \, \mathrm{G}$ and $162.8 \, \mathrm{G}$ and to the immediate left-side slope of the $562.2\, \mathrm{G}$ resonance. Outside of these regions, the \K interaction will be negative, leading to mean-field collapse of the condensate wavefunction. For the discussion that follows, we discard the region around $562.2 \, \mathrm{G}$. We  concentrate on the valley region, where the \K scattering length is strictly positive and widely tunable between $10 \,a_{0}$ and $\infty$. Due to the location of A.1 at the valley border and the appearance of the zero point scattering near the valley center, the interspecies scattering monotonically decreases in the discussed region, and can be set to values $- 62.1 \, a_{0} < a_{\text{NaK}} \leq \infty$. This freedom regarding sign and magnitude of intra- as well as interspecies interaction marks the valley as a rich source for different two-species scenarios, including miscibility, phase separation, collapse and droplet formation. The figure of merit for the identification of different interaction domains is \cite{Miscible1,Miscible2,Miscible3}
\begin{equation} \delta g(B) = \left(g_{\text{NaK}}^{2}(B)	 / \left( g_{\text{Na}} \times g_{\text{K}} (B) \right) \right) - 1, \label{eq:miss} \end{equation}
where $g_{\mathrm{i}}(B) = ((2 \pi \hbar^{2} a_{\mathrm{i}}) / \mu_{\mathrm{i}})$ denotes the interaction parameter for the intra- and interspecies interaction respectively, $\hbar$ being the reduced Planck constant and $\mu_{\mathrm{i}}$ the respective reduced mass. A condensed mixture is said to be miscible if $\delta g < 0$, whereas $\delta g > 0$ implies either immiscibility or mean-field collapse, depending on the sign of $g_{\text{NaK}}$. The inset of Fig.\,\ref{fig5} shows $\delta g(B)$ in the aforementioned magnetic field valley, where four different interaction domains are identified. In the  region $B_{1}:\,B < 109.1 \, \mathrm{G}$, the interspecies interaction outweighs the intraspecies one, $\delta g > 0$, when the atomic wavefunctions will be phase separated, rendering sympathetic cooling of \K inefficient. This changes in the region around the zero point crossing, $B_{2}: 109.1 \, \mathrm{G} < B < 131.5 \, \mathrm{G}$. The clouds are miscible in this domain, but due to the small values of $a_{\text{K}}$ and $a_{\text{NaK}}$, \K can not be efficiently cooled to quantum degeneracy. At $B_{3}: 131.5 \, \mathrm{G} < B < 151.1 \, \mathrm{G}$, both intra- and interspecies interactions are sizable, with $ \delta g > 0$. In this region, two clouds with high density overlap will not be mean-field stable, leading to collapse of the mixture. Finally, at $B_{4}: 151.1 \, \mathrm{G} < B < 162.8\, \mathrm{G}$, $\delta g$ becomes again negative as \K is approaching its second resonance. In this region the individual intraspecies scattering lengths are large enough allowing for efficient evaporation, whereas $a_{\text{NaK}}$ will be small and negative, ensuring miscibility of the cloud. The region $B_{4}$ can therefore be highlighted as most suited for the purpose of condensing both clouds. 

The exactness of Eq.\,\eqref{eq:miss} relies on equal densities, and therefore neither accounts for density imbalances nor includes geometric effects due to the unequal optical trap depths and gravity \cite{ArltMisc}. The mass imbalance $m_{\mathrm{K}}/m_{\mathrm{Na}} \approx 1.7$ partially balances the stronger confinement of $^{39}\mathrm{K}$, giving relative trap frequencies $\omega_{\text{K}}/\omega_{\text{Na}} \approx 1.217$. This leads to a differential gravitational sag of $\delta z_{0}(\omega_{\mathrm{z}}) = (3.19/\omega_{\mathrm{z}}^{2}) \, \mathrm{Hz}^{2}\mathrm{m} $, decreasing the overlap region and shifting the boundaries of the magnetic field domains $B_{1,2,3,4}$. In particular, $B_{4}$ will extend to lower magnetic fields. We lastly point out that at the borders of the collapse region $B_{3}$, beyond mean-field effects can lead to the formation of self-stabilized droplets \cite{Petrov}.
\subsection{Dual-species degeneracy}
\begin{figure}[bth]
\centering
\includegraphics*[width=1\columnwidth]{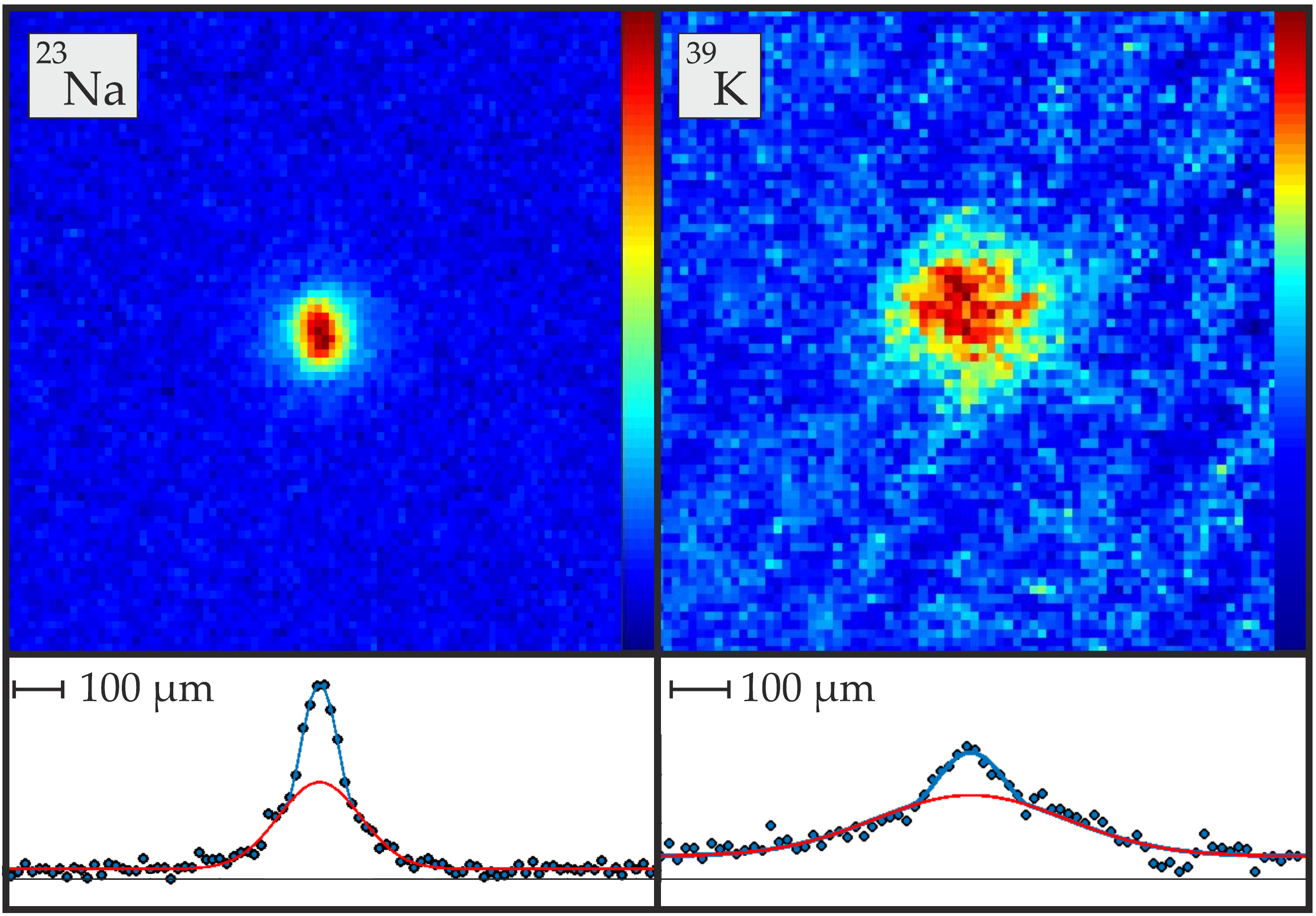}
\caption{Top: Typical absorption images of \Na (left) and \K (right) after a ballistic expansion time of $11.3 \,$ms ($^{23}\mathrm{Na}$) and $14.4\,$ms ($^{23}\mathrm{K}$), respectively. Bottom: Integrated column densities of the top pictures (blue dots) together with bimodal fits of the thermal (red line) and condensed parts (blue line), respectively.}
\label{fig6}
\end{figure}

We realize two stable condensates by operating our cooling procedure at a magnetic field $\tilde{B} = 150.4 \, \mathrm{G}$, where all three scattering lengths are of similar magnitude. In particular, we find $\delta g(\tilde{B}) = 0.047$, when Eq.\,\eqref{eq:miss} would dictate mean-field collapse as $\tilde{B} \in B_{3}$. Due to the previously outlined geometric effect, the reduced cloud overlap can ensure stability at this field strength, depending on the individual density distributions. Following a quick magnetic field ramp to $\tilde{B}$, we perform dual-species evaporation by consecutive linear intensity ramps of the cODT. The final trap frequencies are $\omega_{\text{x,y,z,(Na)}} = 2\pi \: (36,129,157) \, \mathrm{Hz}$. Both species are then released from the optical traps and the bias field is subsequently switched off. To ensure that both clouds are condensed in the same experimental cycle, we successively image both atomic clouds in the same time of flight with a delay of $3\,$ms between the pictures. Figure\,\ref{fig6} shows the resulting density distributions for a typical experimental run as well as the one-dimensional integrated optical densities. A bimodal fit discerns the condensed $(N_{\mathrm{c}})$ and thermal part $(N_{\mathrm{T}})$ of the clouds. For the used parameters, we obtain condensed fractions of $N_{\text{c,Na}} = 42 \, \%$ and $N_{\text{c,K}} = 17 \, \%$ with a total atom number $N_{\mathrm{c}} + N_{\mathrm{T}}$ of $4 \times 10^{4}$ for \Na and $7 \times 10^{4}$ for $^{39}\mathrm{K}$, respectively. 
As outlined before, one possible explanation for the stability of both condensed clouds can be given by the details of our trapping geometry. 
At the final trap frequencies, the differential gravitational sag $\delta z_{0} = 3.27 \, \mathrm{\mu m} $ exceeds the individual (i.e. without interspecies interactions) vertical Thomas-Fermi radii of the condensate wavefunctions, when the condensate wavefunction overlap of the two clouds is already strongly reduced to about $30 \, \%$. Hence the impact of the interspecies interaction is reduced, and mean-field stability can be obtained. We verify this by calculating stationary solutions of the coupled Gross-Pitaevskii equations for the mixture in the Thomas-Fermi approximation. For the used experimental parameters and final observed atom numbers, the derived chemical potentials dictate stability of the condensates. The reduced overlap also modifies the thermalization properties of the mixture. In an interacting and overlapping scenario, both clouds have to be in thermal equilibrium after a thermalization time, hence $T_{\mathrm{Na}} \equiv T_{\mathrm{K}}$. Yet in a forced evaporation sequence with non-interacting or non-overlapping clouds, we will generally have $T_{\mathrm{Na}} \neq T_{\mathrm{K}}$. The intermediate regime, i.e. the effect of reduced interspecies collisions due to decreased spatial overlap, has been studied before \cite{AspectSympathetic}, where a considerable decrease of the interspecies rethermalization rates was found. In our sequence, we determine the individual cloud temperature by Gaussian fits to the wings of the thermal cloud fraction, giving $T_{\mathrm{K}} = 186 \, \mathrm{nK}$ and $T_{\mathrm{Na}} = 87 \, \mathrm{nK}$, respectively. Both values are consistent with the individual extracted condensed fractions and critical temperatures. It can be therefore assumed that in our evaporation sequence, both samples are in thermal equilibrium with themselves, but not with each other. Verifying these findings by numerical simulations demands exact tracing of the evaporation dynamics in presence of the thermal background and are therefore beyond the scope of the stationary Gross-Pitaevskii equation ansatz that was used to support our explanation of mean-field stability. Both effects, the mean-field stability as well as the thermalization properties, are left for future research. 




\section{Conclusion}
We have presented a detailed theoretical and experimental investigation of the \Na and \K collisional properties at ultracold temperatures in the channel $\left|1, -1 \right\rangle_{\text{Na}} \left|1, -1 \right\rangle_{\text{K}}\left|0,0\right\rangle$, in order to achieve dual species condensation. Our theoretical calculations show three distinct signatures, and we have further shown how all of them can be spectrally resolved using atom loss spectroscopy. The measured features coincide well with our theoretical prediction, highlighting the quality of the recent data by \cite{Zwierlein1} and \cite{Bo} on the fermionic isotope pair. The potential energy curves can be further improved by measuring the remaining Feshbach resonances of the bosonic isotope in other collisional channels. We have shown that the interplay of intra- and interspecies scattering lengths gives rise to four different magnetic field regions $B_{\text{1,2,3,4}}$, in which different mixture phenomena can be studied. We have tuned $a_{\mathrm{NaK}}$ to achieve a quantum degenerate mixture by forced optical evaporation. We have further discussed the influence of the differential gravitational sag on the miscibility and the thermalization properties of the mixture. In our system with reduced cloud overlap, we have found the mixture to be out of thermal equilibrium as we switch off the cODT, and the \Na ensemble to be colder due to the lower trap depth. Thermalization can be achieved by raising the trap frequencies after the final stage of forced evaporation, which reduces the differential sag. Combined with the miscible interactions, a high spatial overlap of the atomic clouds gives ideal starting conditions for the association of Feshbach molecules and subsequent molecular ground state spectroscopy. A thorough study of the different interaction regions $B_{\text{1,2,3,4}}$ is left for future work. It is mentioned that the borders of region $B_{3}$ are of considerable interest due to the possibility to study higher-order mean-field effects like the Lee-Huang-Yang correction to the Gross-Pitaevskii equation. This correction was shown to manifest itself in the formation of quantum droplets in regimes which otherwise would display mean-field instabilities. Droplets have been recently shown to emerge in systems of magnetic dipoles \cite{Drop10,Drop1,Drop2} and Zeeman state mixtures \cite{Drop3,Drop4,Drop5}. With the condensed clouds at hand, this region can be revisited and quantum droplets studied. Together with the immiscible region $B_{1}$ and the miscible region $B_{4}$ explored in this article, the magnetic field valley outlined in the inset of Fig.\,\ref{fig5} provides a wealth of possible phase transitions to be studied in future research.
\section{Acknowledgements}
We thank H. Ahlers and M. Siercke for fruitful discussions and P. Gersema for experimental assistance. M. Petzold and E. Schwanke contributed valuably during the early stages of the experiment. We gratefully acknowledge financial support from the European Research Council through ERC Starting Grant POLAR and from the Deutsche Forschungsgemeinschaft (DFG) through CRC 1227 (DQ-mat), project A03 and FOR2247, project E5. K.K.V. thanks the Deutsche Forschungsgemeinschaft for financial support through Research Training Group 1991.

\section*{Appendix: Three-body losses}

The acquired loss data points are fitted to the solution of the coupled differential equation system
\begin{align}\label{Nsodium}
N'_{\mathrm{Na}}=
&-\frac{2}{3} L_{\mathrm{Na},\mathrm{Na},\mathrm{K}} \left<N_{\mathrm{Na}}^2N_\mathrm{K}\right>_V \nonumber\\
&-\frac{1}{3} L_{\mathrm{Na},\mathrm{K},\mathrm{K}} \left<N_{\mathrm{Na}}N_\mathrm{K}^2\right>_V\\
N'_{\mathrm{K}} = 
&- \frac{1}{3}L_{\mathrm{Na},\mathrm{Na},\mathrm{K}} \left<N_{\mathrm{Na}}^2N_\mathrm{K}\right>_V \nonumber\\
&- \frac{2}{3} L_{\mathrm{Na},\mathrm{K},\mathrm{K}} \left<N_{\mathrm{Na}}N_\mathrm{K}^2\right>_V,
\end{align}
using the three-body coefficients $L_{\mathrm{Na},\mathrm{Na},\mathrm{K}}$ and $L_{\mathrm{Na},\mathrm{K},\mathrm{K}}$ as fit parameters. One- and three-body losses due to background collisions heteronuclear collisions are fixed by the corresponding single-species experiments. On the relevant timescale of the experiment, defined here by the effective 1/e lifetime of $^{39}\mathrm{K}$, they appear negligible.
The terms in brackets $\left< ... \right>_V$ refers to the spatially averaged temperature-dependent three-body densities in the volume $V$ for the two different possible loss sources, $\left<N_{\mathrm{A}}^iN_\mathrm{B}^j\right>_V = \int_{V}n^i_{\mathrm{A}} n^j_{\mathrm{B}}d^3r$ where $n_{\mathrm{i}}=n_\mathrm{i}(\mathbf{r},T)$ is the density of species i.
We also consider the effects of anti-evaporation induced heating by solving the additional differential equation
\begin{align}\label{Temp}
3k_BT'\times(N_\mathrm{K}+N_\mathrm{Na}) =
&+ \beta_{\mathrm{Na},\mathrm{Na},\mathrm{K}} L_{\mathrm{Na},\mathrm{Na},\mathrm{K}}  \left<N_{\mathrm{Na}}^2N_\mathrm{K}\right>_V \nonumber\\ 
&+ \beta_{\mathrm{Na},\mathrm{K},\mathrm{K}} L_{\mathrm{Na},\mathrm{K},\mathrm{K}} \left<N_{\mathrm{Na}}N_\mathrm{K}^2\right>_V,
\end{align}
where $k_{B}$ is the Boltzmann constant, together with Eq.\,\eqref{Nsodium}. The quantity $\beta$
\begin{equation} \label{BetaM}
\begin{split}
 \beta_{\mathrm{Na},\mathrm{Na},\mathrm{K}} &=\frac{3}{2}k_BT -  \frac{2}{3}\frac{\int_{V}U_\mathrm{Na} n^2_\mathrm{Na} n_\mathrm{K}d^3r}{\int_{V}n^2_\mathrm{Na} n_\mathrm{K}d^3r}\\
& -\frac{1}{3}\frac{\int_{V}U_\mathrm{K} n_\mathrm{Na} n^2_\mathrm{K}d^3r}{\int_{V}n_\mathrm{Na} n^2_\mathrm{K}d^3r}.
\end{split}
\end{equation}
(and similarly $\beta_{\mathrm{Na},\mathrm{K},\mathrm{K}}$) accounts for the mean potential energy of the lost atoms, with $U$ being the potential energy \cite{Wacker1, efi}. Note that also the second and third term of $\beta$ are temperature dependent through the temperature dependence of $n$.

\bibliography{FeshbachPaper}

\begin{thebibliography}{54}
\expandafter\ifx\csname natexlab\endcsname\relax\def\natexlab#1{#1}\fi
\expandafter\ifx\csname bibnamefont\endcsname\relax
  \def\bibnamefont#1{#1}\fi
\expandafter\ifx\csname bibfnamefont\endcsname\relax
  \def\bibfnamefont#1{#1}\fi
\expandafter\ifx\csname citenamefont\endcsname\relax
  \def\citenamefont#1{#1}\fi
\expandafter\ifx\csname url\endcsname\relax
  \def\url#1{\texttt{#1}}\fi
\expandafter\ifx\csname urlprefix\endcsname\relax\def\urlprefix{URL }\fi
\providecommand{\bibinfo}[2]{#2}
\providecommand{\eprint}[2][]{\url{#2}}

\bibitem[{\citenamefont{D'Incao et~al.}(2017)\citenamefont{D'Incao, Krutzik,
  Elliott, and Williams}}]{CAL}
\bibinfo{author}{\bibfnamefont{J.~P.} \bibnamefont{D'Incao}},
  \bibinfo{author}{\bibfnamefont{M.}~\bibnamefont{Krutzik}},
  \bibinfo{author}{\bibfnamefont{E.}~\bibnamefont{Elliott}}, \bibnamefont{and}
  \bibinfo{author}{\bibfnamefont{J.~R.} \bibnamefont{Williams}},
  \bibinfo{journal}{Phys. Rev. A} \textbf{\bibinfo{volume}{95}},
  \bibinfo{pages}{012701} (\bibinfo{year}{2017}),
  \urlprefix\url{https://link.aps.org/doi/10.1103/PhysRevA.95.012701}.

\bibitem[{\citenamefont{Lin et~al.}(2011)\citenamefont{Lin, Jiménez-García, and
  Spielman}}]{SOC}
\bibinfo{author}{\bibfnamefont{Y.-J.} \bibnamefont{Lin}},
  \bibinfo{author}{\bibfnamefont{K.}~\bibnamefont{Jiménez-García}},
  \bibnamefont{and} \bibinfo{author}{\bibfnamefont{I.~B.}
  \bibnamefont{Spielman}}, \bibinfo{journal}{Nature 471, 83–86}
  (\bibinfo{year}{2011}).

\bibitem[{\citenamefont{Pollet et~al.}(2010)\citenamefont{Pollet, Picon,
  B\"uchler, and Troyer}}]{Buechler}
\bibinfo{author}{\bibfnamefont{L.}~\bibnamefont{Pollet}},
  \bibinfo{author}{\bibfnamefont{J.~D.} \bibnamefont{Picon}},
  \bibinfo{author}{\bibfnamefont{H.~P.} \bibnamefont{B\"uchler}},
  \bibnamefont{and} \bibinfo{author}{\bibfnamefont{M.}~\bibnamefont{Troyer}},
  \bibinfo{journal}{Phys. Rev. Lett.} \textbf{\bibinfo{volume}{104}},
  \bibinfo{pages}{125302} (\bibinfo{year}{2010}),
  \urlprefix\url{https://link.aps.org/doi/10.1103/PhysRevLett.104.125302}.

\bibitem[{\citenamefont{Lahaye et~al.}(2009)\citenamefont{Lahaye, Menotti,
  Santos, Lewenstein, and Pfau}}]{Lahaye}
\bibinfo{author}{\bibfnamefont{T.}~\bibnamefont{Lahaye}},
  \bibinfo{author}{\bibfnamefont{C.}~\bibnamefont{Menotti}},
  \bibinfo{author}{\bibfnamefont{L.}~\bibnamefont{Santos}},
  \bibinfo{author}{\bibfnamefont{M.}~\bibnamefont{Lewenstein}},
  \bibnamefont{and} \bibinfo{author}{\bibfnamefont{T.}~\bibnamefont{Pfau}},
  \bibinfo{journal}{Reports on Progress in Physics}
  \textbf{\bibinfo{volume}{72}}, \bibinfo{pages}{126401}
  (\bibinfo{year}{2009}),
  \urlprefix\url{http://stacks.iop.org/0034-4885/72/i=12/a=126401}.

\bibitem[{\citenamefont{Ni et~al.}(2008)\citenamefont{Ni, Ospelkaus,
  de~Miranda, Peer, Neyenhuis, Zirbel, Kotochigova, Julienne, Jin, and
  Ye}}]{KRb1}
\bibinfo{author}{\bibfnamefont{K.-K.} \bibnamefont{Ni}},
  \bibinfo{author}{\bibfnamefont{S.}~\bibnamefont{Ospelkaus}},
  \bibinfo{author}{\bibfnamefont{M.~H.~G.} \bibnamefont{de~Miranda}},
  \bibinfo{author}{\bibfnamefont{A.}~\bibnamefont{Peer}},
  \bibinfo{author}{\bibfnamefont{B.}~\bibnamefont{Neyenhuis}},
  \bibinfo{author}{\bibfnamefont{J.~J.} \bibnamefont{Zirbel}},
  \bibinfo{author}{\bibfnamefont{S.}~\bibnamefont{Kotochigova}},
  \bibinfo{author}{\bibfnamefont{P.~S.} \bibnamefont{Julienne}},
  \bibinfo{author}{\bibfnamefont{D.~S.} \bibnamefont{Jin}}, \bibnamefont{and}
  \bibinfo{author}{\bibfnamefont{J.}~\bibnamefont{Ye}},
  \bibinfo{journal}{Science} \textbf{\bibinfo{volume}{322}},
  \bibinfo{pages}{231} (\bibinfo{year}{2008}).

\bibitem[{\citenamefont{Ospelkaus et~al.}(2009)\citenamefont{Ospelkaus, Ni,
  de~Miranda, Neyenhuis, Wang, Kotochigova, Julienne, Jin, and Ye}}]{KRb2}
\bibinfo{author}{\bibfnamefont{S.}~\bibnamefont{Ospelkaus}},
  \bibinfo{author}{\bibfnamefont{K.-K.} \bibnamefont{Ni}},
  \bibinfo{author}{\bibfnamefont{M.~H.~G.} \bibnamefont{de~Miranda}},
  \bibinfo{author}{\bibfnamefont{B.}~\bibnamefont{Neyenhuis}},
  \bibinfo{author}{\bibfnamefont{D.}~\bibnamefont{Wang}},
  \bibinfo{author}{\bibfnamefont{S.}~\bibnamefont{Kotochigova}},
  \bibinfo{author}{\bibfnamefont{P.~S.} \bibnamefont{Julienne}},
  \bibinfo{author}{\bibfnamefont{D.~S.} \bibnamefont{Jin}}, \bibnamefont{and}
  \bibinfo{author}{\bibfnamefont{J.}~\bibnamefont{Ye}},
  \bibinfo{journal}{Faraday Discuss.} \textbf{\bibinfo{volume}{142}},
  \bibinfo{pages}{351} (\bibinfo{year}{2009}),
  \urlprefix\url{http://dx.doi.org/10.1039/B821298H}.

\bibitem[{\citenamefont{S.Ospelkaus et~al.}(2010)\citenamefont{S.Ospelkaus, Ni,
  Wang, de~Miranda, Neyenhuis, Quemener, Julienne, Bohn, Jin, and
  Ye}}]{Ospelkaus2010}
\bibinfo{author}{\bibnamefont{S.Ospelkaus}},
  \bibinfo{author}{\bibfnamefont{K.-K.} \bibnamefont{Ni}},
  \bibinfo{author}{\bibfnamefont{D.}~\bibnamefont{Wang}},
  \bibinfo{author}{\bibfnamefont{M.~H.~G.} \bibnamefont{de~Miranda}},
  \bibinfo{author}{\bibfnamefont{B.}~\bibnamefont{Neyenhuis}},
  \bibinfo{author}{\bibfnamefont{G.}~\bibnamefont{Quemener}},
  \bibinfo{author}{\bibfnamefont{P.~S.} \bibnamefont{Julienne}},
  \bibinfo{author}{\bibfnamefont{J.~L.} \bibnamefont{Bohn}},
  \bibinfo{author}{\bibfnamefont{D.~S.} \bibnamefont{Jin}}, \bibnamefont{and}
  \bibinfo{author}{\bibfnamefont{J.}~\bibnamefont{Ye}},
  \bibinfo{journal}{Science} \textbf{\bibinfo{volume}{327}},
  \bibinfo{pages}{853} (\bibinfo{year}{2010}).

\bibitem[{\citenamefont{Heo et~al.}(2012)\citenamefont{Heo, Wang, Christensen,
  Rvachov, Cotta, Choi, Lee, and Ketterle}}]{LiNa}
\bibinfo{author}{\bibfnamefont{M.-S.} \bibnamefont{Heo}},
  \bibinfo{author}{\bibfnamefont{T.~T.} \bibnamefont{Wang}},
  \bibinfo{author}{\bibfnamefont{C.~A.} \bibnamefont{Christensen}},
  \bibinfo{author}{\bibfnamefont{T.~M.} \bibnamefont{Rvachov}},
  \bibinfo{author}{\bibfnamefont{D.~A.} \bibnamefont{Cotta}},
  \bibinfo{author}{\bibfnamefont{J.-H.} \bibnamefont{Choi}},
  \bibinfo{author}{\bibfnamefont{Y.-R.} \bibnamefont{Lee}}, \bibnamefont{and}
  \bibinfo{author}{\bibfnamefont{W.}~\bibnamefont{Ketterle}},
  \bibinfo{journal}{Phys. Rev. A} \textbf{\bibinfo{volume}{86}},
  \bibinfo{pages}{021602} (\bibinfo{year}{2012}),
  \urlprefix\url{https://link.aps.org/doi/10.1103/PhysRevA.86.021602}.

\bibitem[{\citenamefont{Reichs\"ollner
  et~al.}(2017)\citenamefont{Reichs\"ollner, Schindewolf, Takekoshi, Grimm, and
  N\"agerl}}]{RbCs}
\bibinfo{author}{\bibfnamefont{L.}~\bibnamefont{Reichs\"ollner}},
  \bibinfo{author}{\bibfnamefont{A.}~\bibnamefont{Schindewolf}},
  \bibinfo{author}{\bibfnamefont{T.}~\bibnamefont{Takekoshi}},
  \bibinfo{author}{\bibfnamefont{R.}~\bibnamefont{Grimm}}, \bibnamefont{and}
  \bibinfo{author}{\bibfnamefont{H.-C.} \bibnamefont{N\"agerl}},
  \bibinfo{journal}{Phys. Rev. Lett.} \textbf{\bibinfo{volume}{118}},
  \bibinfo{pages}{073201} (\bibinfo{year}{2017}),
  \urlprefix\url{https://link.aps.org/doi/10.1103/PhysRevLett.118.073201}.

\bibitem[{\citenamefont{Guo et~al.}(2016)\citenamefont{Guo, Zhu, Lu, Ye, Wang,
  Vexiau, Bouloufa-Maafa, Qu\'em\'ener, Dulieu, and Wang}}]{NaRbREVISION}
\bibinfo{author}{\bibfnamefont{M.}~\bibnamefont{Guo}},
  \bibinfo{author}{\bibfnamefont{B.}~\bibnamefont{Zhu}},
  \bibinfo{author}{\bibfnamefont{B.}~\bibnamefont{Lu}},
  \bibinfo{author}{\bibfnamefont{X.}~\bibnamefont{Ye}},
  \bibinfo{author}{\bibfnamefont{F.}~\bibnamefont{Wang}},
  \bibinfo{author}{\bibfnamefont{R.}~\bibnamefont{Vexiau}},
  \bibinfo{author}{\bibfnamefont{N.}~\bibnamefont{Bouloufa-Maafa}},
  \bibinfo{author}{\bibfnamefont{G.}~\bibnamefont{Qu\'em\'ener}},
  \bibinfo{author}{\bibfnamefont{O.}~\bibnamefont{Dulieu}}, \bibnamefont{and}
  \bibinfo{author}{\bibfnamefont{D.}~\bibnamefont{Wang}},
  \bibinfo{journal}{Phys. Rev. Lett.} \textbf{\bibinfo{volume}{116}},
  \bibinfo{pages}{205303} (\bibinfo{year}{2016}),
  \urlprefix\url{https://link.aps.org/doi/10.1103/PhysRevLett.116.205303}.

\bibitem[{\citenamefont{H\"afner et~al.}(2017)\citenamefont{H\"afner, Ulmanis,
  Kuhnle, Wang, Greene, and Weidem\"uller}}]{LiCs}
\bibinfo{author}{\bibfnamefont{S.}~\bibnamefont{H\"afner}},
  \bibinfo{author}{\bibfnamefont{J.}~\bibnamefont{Ulmanis}},
  \bibinfo{author}{\bibfnamefont{E.~D.} \bibnamefont{Kuhnle}},
  \bibinfo{author}{\bibfnamefont{Y.}~\bibnamefont{Wang}},
  \bibinfo{author}{\bibfnamefont{C.~H.} \bibnamefont{Greene}},
  \bibnamefont{and}
  \bibinfo{author}{\bibfnamefont{M.}~\bibnamefont{Weidem\"uller}},
  \bibinfo{journal}{Phys. Rev. A} \textbf{\bibinfo{volume}{95}},
  \bibinfo{pages}{062708} (\bibinfo{year}{2017}),
  \urlprefix\url{https://link.aps.org/doi/10.1103/PhysRevA.95.062708}.

\bibitem[{\citenamefont{Vaidya et~al.}(2015)\citenamefont{Vaidya, Tiamsuphat,
  Rolston, and Porto}}]{RbYb}
\bibinfo{author}{\bibfnamefont{V.~D.} \bibnamefont{Vaidya}},
  \bibinfo{author}{\bibfnamefont{J.}~\bibnamefont{Tiamsuphat}},
  \bibinfo{author}{\bibfnamefont{S.~L.} \bibnamefont{Rolston}},
  \bibnamefont{and} \bibinfo{author}{\bibfnamefont{J.~V.} \bibnamefont{Porto}},
  \bibinfo{journal}{Phys. Rev. A} \textbf{\bibinfo{volume}{92}},
  \bibinfo{pages}{043604} (\bibinfo{year}{2015}),
  \urlprefix\url{https://link.aps.org/doi/10.1103/PhysRevA.92.043604}.

\bibitem[{\citenamefont{Barratt}(1924)}]{Barratt1924}
\bibinfo{author}{\bibfnamefont{S.}~\bibnamefont{Barratt}},
  \bibinfo{journal}{Proceedings of the Royal Society A: Mathematical, Physical
  and Engineering Sciences} \textbf{\bibinfo{volume}{105}},
  \bibinfo{pages}{221} (\bibinfo{year}{1924}).

\bibitem[{\citenamefont{Ross et~al.}(1986)\citenamefont{Ross, Effantin,
  d'Incan, and Barrow}}]{Barrow}
\bibinfo{author}{\bibfnamefont{A.~J.} \bibnamefont{Ross}},
  \bibinfo{author}{\bibfnamefont{C.}~\bibnamefont{Effantin}},
  \bibinfo{author}{\bibfnamefont{J.}~\bibnamefont{d'Incan}}, \bibnamefont{and}
  \bibinfo{author}{\bibfnamefont{R.~F.} \bibnamefont{Barrow}},
  \bibinfo{journal}{J. Phys. B: At., Mol. Phys.} \textbf{\bibinfo{volume}{19}},
  \bibinfo{pages}{1449} (\bibinfo{year}{1986}).

\bibitem[{\citenamefont{Kasahara et~al.}(1991)\citenamefont{Kasahara, Baba, and
  Kat{$\mathrm{\hat{o}}$}}}]{Kato}
\bibinfo{author}{\bibfnamefont{S.}~\bibnamefont{Kasahara}},
  \bibinfo{author}{\bibfnamefont{M.}~\bibnamefont{Baba}}, \bibnamefont{and}
  \bibinfo{author}{\bibfnamefont{H.}~\bibnamefont{Kat{$\mathrm{\hat{o}}$}}},
  \bibinfo{journal}{J. Chem. Phys.} \textbf{\bibinfo{volume}{94}},
  \bibinfo{pages}{7713} (\bibinfo{year}{1991}).

\bibitem[{\citenamefont{Zemke and Stwalley}(1999)}]{Zemke}
\bibinfo{author}{\bibfnamefont{W.}~\bibnamefont{Zemke}} \bibnamefont{and}
  \bibinfo{author}{\bibfnamefont{W.}~\bibnamefont{Stwalley}},
  \bibinfo{journal}{J. Chem. Phys.} \textbf{\bibinfo{volume}{111}},
  \bibinfo{pages}{4956} (\bibinfo{year}{1999}).

\bibitem[{\citenamefont{Ferber et~al.}(2000)\citenamefont{Ferber, Pazyuk,
  Stolyarov, Zaitsevskii, Kowalczyk, Chen, Wang, and Stwalley}}]{Stolyarov}
\bibinfo{author}{\bibfnamefont{R.}~\bibnamefont{Ferber}},
  \bibinfo{author}{\bibfnamefont{E.~A.} \bibnamefont{Pazyuk}},
  \bibinfo{author}{\bibfnamefont{A.~V.} \bibnamefont{Stolyarov}},
  \bibinfo{author}{\bibfnamefont{A.}~\bibnamefont{Zaitsevskii}},
  \bibinfo{author}{\bibfnamefont{P.}~\bibnamefont{Kowalczyk}},
  \bibinfo{author}{\bibfnamefont{H.}~\bibnamefont{Chen}},
  \bibinfo{author}{\bibfnamefont{H.}~\bibnamefont{Wang}}, \bibnamefont{and}
  \bibinfo{author}{\bibfnamefont{W.~C.} \bibnamefont{Stwalley}},
  \bibinfo{journal}{J. Chem. Phys.} \textbf{\bibinfo{volume}{112}},
  \bibinfo{pages}{5740} (\bibinfo{year}{2000}).

\bibitem[{\citenamefont{Russier-Antoine
  et~al.}(2000)\citenamefont{Russier-Antoine, Ross, Aubert-Frecon, Martin, and
  Crozet}}]{Russier}
\bibinfo{author}{\bibfnamefont{I.}~\bibnamefont{Russier-Antoine}},
  \bibinfo{author}{\bibfnamefont{A.}~\bibnamefont{Ross}},
  \bibinfo{author}{\bibfnamefont{M.}~\bibnamefont{Aubert-Frecon}},
  \bibinfo{author}{\bibfnamefont{F.}~\bibnamefont{Martin}}, \bibnamefont{and}
  \bibinfo{author}{\bibfnamefont{P.}~\bibnamefont{Crozet}},
  \bibinfo{journal}{J. Phys. B: At., Mol. Phys.} \textbf{\bibinfo{volume}{33}},
  \bibinfo{pages}{2753} (\bibinfo{year}{2000}).

\bibitem[{\citenamefont{Gerdes et~al.}(2008)\citenamefont{Gerdes, Hobein,
  Kn{\"o}ckel, and Tiemann}}]{Gerdes2008}
\bibinfo{author}{\bibfnamefont{A.}~\bibnamefont{Gerdes}},
  \bibinfo{author}{\bibfnamefont{M.}~\bibnamefont{Hobein}},
  \bibinfo{author}{\bibfnamefont{H.}~\bibnamefont{Kn{\"o}ckel}},
  \bibnamefont{and} \bibinfo{author}{\bibfnamefont{E.}~\bibnamefont{Tiemann}},
  \bibinfo{journal}{Eur. Phys. J. D} \textbf{\bibinfo{volume}{49}},
  \bibinfo{pages}{67} (\bibinfo{year}{2008}).

\bibitem[{\citenamefont{Park et~al.}(2012)\citenamefont{Park, Wu, Santiago,
  Tiecke, Will, Ahmadi, and Zwierlein}}]{Zwierlein1}
\bibinfo{author}{\bibfnamefont{J.~W.} \bibnamefont{Park}},
  \bibinfo{author}{\bibfnamefont{C.-H.} \bibnamefont{Wu}},
  \bibinfo{author}{\bibfnamefont{I.}~\bibnamefont{Santiago}},
  \bibinfo{author}{\bibfnamefont{T.~G.} \bibnamefont{Tiecke}},
  \bibinfo{author}{\bibfnamefont{S.}~\bibnamefont{Will}},
  \bibinfo{author}{\bibfnamefont{P.}~\bibnamefont{Ahmadi}}, \bibnamefont{and}
  \bibinfo{author}{\bibfnamefont{M.~W.} \bibnamefont{Zwierlein}},
  \bibinfo{journal}{Phys. Rev. A} \textbf{\bibinfo{volume}{85}},
  \bibinfo{pages}{051602} (\bibinfo{year}{2012}),
  \urlprefix\url{https://link.aps.org/doi/10.1103/PhysRevA.85.051602}.

\bibitem[{\citenamefont{Wu et~al.}(2012)\citenamefont{Wu, Park, Ahmadi, Will,
  and Zwierlein}}]{Zwierlein2}
\bibinfo{author}{\bibfnamefont{C.-H.} \bibnamefont{Wu}},
  \bibinfo{author}{\bibfnamefont{J.~W.} \bibnamefont{Park}},
  \bibinfo{author}{\bibfnamefont{P.}~\bibnamefont{Ahmadi}},
  \bibinfo{author}{\bibfnamefont{S.}~\bibnamefont{Will}}, \bibnamefont{and}
  \bibinfo{author}{\bibfnamefont{M.~W.} \bibnamefont{Zwierlein}},
  \bibinfo{journal}{Phys. Rev. Lett.} \textbf{\bibinfo{volume}{109}},
  \bibinfo{pages}{085301} (\bibinfo{year}{2012}).

\bibitem[{\citenamefont{Park et~al.}(2015)\citenamefont{Park, Will, and
  Zwierlein}}]{Zwierlein3}
\bibinfo{author}{\bibfnamefont{J.~W.} \bibnamefont{Park}},
  \bibinfo{author}{\bibfnamefont{S.~A.} \bibnamefont{Will}}, \bibnamefont{and}
  \bibinfo{author}{\bibfnamefont{M.~W.} \bibnamefont{Zwierlein}},
  \bibinfo{journal}{Phys. Rev. Lett.} \textbf{\bibinfo{volume}{114}},
  \bibinfo{pages}{205302} (\bibinfo{year}{2015}),
  \urlprefix\url{https://link.aps.org/doi/10.1103/PhysRevLett.114.205302}.

\bibitem[{\citenamefont{Zhu et~al.}(2017)\citenamefont{Zhu, Yang, Liu, Zhang,
  Liu, Nan, Rui, Zhao, Pan, and Tiemann}}]{Bo}
\bibinfo{author}{\bibfnamefont{M.-J.} \bibnamefont{Zhu}},
  \bibinfo{author}{\bibfnamefont{H.}~\bibnamefont{Yang}},
  \bibinfo{author}{\bibfnamefont{L.}~\bibnamefont{Liu}},
  \bibinfo{author}{\bibfnamefont{D.-C.} \bibnamefont{Zhang}},
  \bibinfo{author}{\bibfnamefont{Y.-X.} \bibnamefont{Liu}},
  \bibinfo{author}{\bibfnamefont{J.}~\bibnamefont{Nan}},
  \bibinfo{author}{\bibfnamefont{J.}~\bibnamefont{Rui}},
  \bibinfo{author}{\bibfnamefont{B.}~\bibnamefont{Zhao}},
  \bibinfo{author}{\bibfnamefont{J.-W.} \bibnamefont{Pan}}, \bibnamefont{and}
  \bibinfo{author}{\bibfnamefont{E.}~\bibnamefont{Tiemann}},
  \bibinfo{journal}{Physical Review A} \textbf{\bibinfo{volume}{96}}
  (\bibinfo{year}{2017}).

\bibitem[{\citenamefont{See{\ss}elberg
  et~al.}(2018)\citenamefont{See{\ss}elberg, Buchheim, Lu, Schneider, Luo,
  Tiemann, Bloch, and Gohle}}]{Frauke}
\bibinfo{author}{\bibfnamefont{F.}~\bibnamefont{See{\ss}elberg}},
  \bibinfo{author}{\bibfnamefont{N.}~\bibnamefont{Buchheim}},
  \bibinfo{author}{\bibfnamefont{Z.-K.} \bibnamefont{Lu}},
  \bibinfo{author}{\bibfnamefont{T.}~\bibnamefont{Schneider}},
  \bibinfo{author}{\bibfnamefont{X.-Y.} \bibnamefont{Luo}},
  \bibinfo{author}{\bibfnamefont{E.}~\bibnamefont{Tiemann}},
  \bibinfo{author}{\bibfnamefont{I.}~\bibnamefont{Bloch}}, \bibnamefont{and}
  \bibinfo{author}{\bibfnamefont{C.}~\bibnamefont{Gohle}},
  \bibinfo{journal}{Physical Review A} \textbf{\bibinfo{volume}{97}}
  (\bibinfo{year}{2018}).

\bibitem[{\citenamefont{Temelkov et~al.}(2015)\citenamefont{Temelkov,
  Kn\"ockel, Pashov, and Tiemann}}]{Temelkov}
\bibinfo{author}{\bibfnamefont{I.}~\bibnamefont{Temelkov}},
  \bibinfo{author}{\bibfnamefont{H.}~\bibnamefont{Kn\"ockel}},
  \bibinfo{author}{\bibfnamefont{A.}~\bibnamefont{Pashov}}, \bibnamefont{and}
  \bibinfo{author}{\bibfnamefont{E.}~\bibnamefont{Tiemann}},
  \bibinfo{journal}{Phys. Rev. A} \textbf{\bibinfo{volume}{91}},
  \bibinfo{pages}{032512} (\bibinfo{year}{2015}),
  \urlprefix\url{https://link.aps.org/doi/10.1103/PhysRevA.91.032512}.

\bibitem[{\citenamefont{Viel and Simoni}(2016)}]{VielSimoni}
\bibinfo{author}{\bibfnamefont{A.}~\bibnamefont{Viel}} \bibnamefont{and}
  \bibinfo{author}{\bibfnamefont{A.}~\bibnamefont{Simoni}},
  \bibinfo{journal}{Phys. Rev. A} \textbf{\bibinfo{volume}{93}},
  \bibinfo{pages}{042701} (\bibinfo{year}{2016}),
  \urlprefix\url{https://link.aps.org/doi/10.1103/PhysRevA.93.042701}.

\bibitem[{\citenamefont{Gempel}(2016)}]{Gempel}
\bibinfo{author}{\bibfnamefont{M.~W.} \bibnamefont{Gempel}}, Ph.D. thesis
  (\bibinfo{year}{2016}).

\bibitem[{\citenamefont{Landini et~al.}(2011)\citenamefont{Landini, Roy,
  Carcagn\'{\i}, Trypogeorgos, Fattori, Inguscio, and Modugno}}]{Landini}
\bibinfo{author}{\bibfnamefont{M.}~\bibnamefont{Landini}},
  \bibinfo{author}{\bibfnamefont{S.}~\bibnamefont{Roy}},
  \bibinfo{author}{\bibfnamefont{L.}~\bibnamefont{Carcagn\'{\i}}},
  \bibinfo{author}{\bibfnamefont{D.}~\bibnamefont{Trypogeorgos}},
  \bibinfo{author}{\bibfnamefont{M.}~\bibnamefont{Fattori}},
  \bibinfo{author}{\bibfnamefont{M.}~\bibnamefont{Inguscio}}, \bibnamefont{and}
  \bibinfo{author}{\bibfnamefont{G.}~\bibnamefont{Modugno}},
  \bibinfo{journal}{Phys. Rev. A} \textbf{\bibinfo{volume}{84}},
  \bibinfo{pages}{043432} (\bibinfo{year}{2011}),
  \urlprefix\url{https://link.aps.org/doi/10.1103/PhysRevA.84.043432}.

\bibitem[{\citenamefont{Weber et~al.}(2003)\citenamefont{Weber, Herbig, Mark,
  N\"agerl, and Grimm}}]{efi}
\bibinfo{author}{\bibfnamefont{T.}~\bibnamefont{Weber}},
  \bibinfo{author}{\bibfnamefont{J.}~\bibnamefont{Herbig}},
  \bibinfo{author}{\bibfnamefont{M.}~\bibnamefont{Mark}},
  \bibinfo{author}{\bibfnamefont{H.-C.} \bibnamefont{N\"agerl}},
  \bibnamefont{and} \bibinfo{author}{\bibfnamefont{R.}~\bibnamefont{Grimm}},
  \bibinfo{journal}{Phys. Rev. Lett.} \textbf{\bibinfo{volume}{91}},
  \bibinfo{pages}{123201} (\bibinfo{year}{2003}),
  \urlprefix\url{https://link.aps.org/doi/10.1103/PhysRevLett.91.123201}.

\bibitem[{\citenamefont{Wacker et~al.}(2016)\citenamefont{Wacker, J\o{}rgensen,
  Birkmose, Winter, Mikkelsen, Sherson, Zinner, and Arlt}}]{Wacker1}
\bibinfo{author}{\bibfnamefont{L.~J.} \bibnamefont{Wacker}},
  \bibinfo{author}{\bibfnamefont{N.~B.} \bibnamefont{J\o{}rgensen}},
  \bibinfo{author}{\bibfnamefont{D.}~\bibnamefont{Birkmose}},
  \bibinfo{author}{\bibfnamefont{N.}~\bibnamefont{Winter}},
  \bibinfo{author}{\bibfnamefont{M.}~\bibnamefont{Mikkelsen}},
  \bibinfo{author}{\bibfnamefont{J.}~\bibnamefont{Sherson}},
  \bibinfo{author}{\bibfnamefont{N.}~\bibnamefont{Zinner}}, \bibnamefont{and}
  \bibinfo{author}{\bibfnamefont{J.~J.} \bibnamefont{Arlt}},
  \bibinfo{journal}{Phys. Rev. Lett.} \textbf{\bibinfo{volume}{117}},
  \bibinfo{pages}{163201} (\bibinfo{year}{2016}),
  \urlprefix\url{https://link.aps.org/doi/10.1103/PhysRevLett.117.163201}.

\bibitem[{\citenamefont{Efimov}(1970)}]{Efimov1970ela}
\bibinfo{author}{\bibfnamefont{V.}~\bibnamefont{Efimov}},
  \bibinfo{journal}{Phys. Lett. B} \textbf{\bibinfo{volume}{33}},
  \bibinfo{pages}{563} (\bibinfo{year}{1970}).

\bibitem[{\citenamefont{Kraemer et~al.}(2006)\citenamefont{Kraemer, Mark,
  Waldburger, Danzl, Chin, Engeser, Lange, Pilch, Jaakkola, N\"agerl
  et~al.}}]{Kraemer2006efe}
\bibinfo{author}{\bibfnamefont{T.}~\bibnamefont{Kraemer}},
  \bibinfo{author}{\bibfnamefont{M.}~\bibnamefont{Mark}},
  \bibinfo{author}{\bibfnamefont{P.}~\bibnamefont{Waldburger}},
  \bibinfo{author}{\bibfnamefont{J.~G.} \bibnamefont{Danzl}},
  \bibinfo{author}{\bibfnamefont{C.}~\bibnamefont{Chin}},
  \bibinfo{author}{\bibfnamefont{B.}~\bibnamefont{Engeser}},
  \bibinfo{author}{\bibfnamefont{A.~D.} \bibnamefont{Lange}},
  \bibinfo{author}{\bibfnamefont{K.}~\bibnamefont{Pilch}},
  \bibinfo{author}{\bibfnamefont{A.}~\bibnamefont{Jaakkola}},
  \bibinfo{author}{\bibfnamefont{H.-C.} \bibnamefont{N\"agerl}},
  \bibnamefont{et~al.}, \bibinfo{journal}{Nature}
  \textbf{\bibinfo{volume}{440}}, \bibinfo{pages}{315} (\bibinfo{year}{2006}).

\bibitem[{\citenamefont{Bloom et~al.}(2013)\citenamefont{Bloom, Hu, Cumby, and
  Jin}}]{bloom2013}
\bibinfo{author}{\bibfnamefont{R.~S.} \bibnamefont{Bloom}},
  \bibinfo{author}{\bibfnamefont{M.-G.} \bibnamefont{Hu}},
  \bibinfo{author}{\bibfnamefont{T.~D.} \bibnamefont{Cumby}}, \bibnamefont{and}
  \bibinfo{author}{\bibfnamefont{D.~S.} \bibnamefont{Jin}},
  \bibinfo{journal}{Phys. Rev. Lett.} \textbf{\bibinfo{volume}{111}},
  \bibinfo{pages}{105301} (\bibinfo{year}{2013}).

\bibitem[{\citenamefont{Pires et~al.}(2014)\citenamefont{Pires, Ulmanis,
  H\"afner, Repp, Arias, Kuhnle, and Weidem\"uller}}]{Pires2014}
\bibinfo{author}{\bibfnamefont{R.}~\bibnamefont{Pires}},
  \bibinfo{author}{\bibfnamefont{J.}~\bibnamefont{Ulmanis}},
  \bibinfo{author}{\bibfnamefont{S.}~\bibnamefont{H\"afner}},
  \bibinfo{author}{\bibfnamefont{M.}~\bibnamefont{Repp}},
  \bibinfo{author}{\bibfnamefont{A.}~\bibnamefont{Arias}},
  \bibinfo{author}{\bibfnamefont{E.~D.} \bibnamefont{Kuhnle}},
  \bibnamefont{and}
  \bibinfo{author}{\bibfnamefont{M.}~\bibnamefont{Weidem\"uller}},
  \bibinfo{journal}{Phys. Rev. Lett.} \textbf{\bibinfo{volume}{112}},
  \bibinfo{pages}{250404} (\bibinfo{year}{2014}).

\bibitem[{\citenamefont{Tung et~al.}(2014)\citenamefont{Tung,
  Jim\'enez-Garc\'{\i}a, Johansen, Parker, and Chin}}]{Tung2014}
\bibinfo{author}{\bibfnamefont{S.-K.} \bibnamefont{Tung}},
  \bibinfo{author}{\bibfnamefont{K.}~\bibnamefont{Jim\'enez-Garc\'{\i}a}},
  \bibinfo{author}{\bibfnamefont{J.}~\bibnamefont{Johansen}},
  \bibinfo{author}{\bibfnamefont{C.~V.} \bibnamefont{Parker}},
  \bibnamefont{and} \bibinfo{author}{\bibfnamefont{C.}~\bibnamefont{Chin}},
  \bibinfo{journal}{Phys. Rev. Lett.} \textbf{\bibinfo{volume}{113}},
  \bibinfo{pages}{240402} (\bibinfo{year}{2014}).

\bibitem[{\citenamefont{Stenger et~al.}(1999)\citenamefont{Stenger, Inouye,
  Andrews, Miesner, Stamper-Kurn, and Ketterle}}]{Ketterle}
\bibinfo{author}{\bibfnamefont{J.}~\bibnamefont{Stenger}},
  \bibinfo{author}{\bibfnamefont{S.}~\bibnamefont{Inouye}},
  \bibinfo{author}{\bibfnamefont{M.~R.} \bibnamefont{Andrews}},
  \bibinfo{author}{\bibfnamefont{H.-J.} \bibnamefont{Miesner}},
  \bibinfo{author}{\bibfnamefont{D.~M.} \bibnamefont{Stamper-Kurn}},
  \bibnamefont{and} \bibinfo{author}{\bibfnamefont{W.}~\bibnamefont{Ketterle}},
  \bibinfo{journal}{Phys. Rev. Lett.} \textbf{\bibinfo{volume}{82}},
  \bibinfo{pages}{2422} (\bibinfo{year}{1999}),
  \urlprefix\url{https://link.aps.org/doi/10.1103/PhysRevLett.82.2422}.

\bibitem[{\citenamefont{D'Errico et~al.}(2007)\citenamefont{D'Errico, Zaccanti,
  Fattori, Roati, Inguscio, Modugno, and Simoni}}]{K39feshbach}
\bibinfo{author}{\bibfnamefont{C.}~\bibnamefont{D'Errico}},
  \bibinfo{author}{\bibfnamefont{M.}~\bibnamefont{Zaccanti}},
  \bibinfo{author}{\bibfnamefont{M.}~\bibnamefont{Fattori}},
  \bibinfo{author}{\bibfnamefont{G.}~\bibnamefont{Roati}},
  \bibinfo{author}{\bibfnamefont{M.}~\bibnamefont{Inguscio}},
  \bibinfo{author}{\bibfnamefont{G.}~\bibnamefont{Modugno}}, \bibnamefont{and}
  \bibinfo{author}{\bibfnamefont{A.}~\bibnamefont{Simoni}},
  \bibinfo{journal}{New Journal of Physics} \textbf{\bibinfo{volume}{9}},
  \bibinfo{pages}{223} (\bibinfo{year}{2007}),
  \urlprefix\url{http://stacks.iop.org/1367-2630/9/i=7/a=223}.

\bibitem[{\citenamefont{Salomon et~al.}(2014)\citenamefont{Salomon, Fouch\'e,
  Lepoutre, Aspect, and Bourdel}}]{AspectK39}
\bibinfo{author}{\bibfnamefont{G.}~\bibnamefont{Salomon}},
  \bibinfo{author}{\bibfnamefont{L.}~\bibnamefont{Fouch\'e}},
  \bibinfo{author}{\bibfnamefont{S.}~\bibnamefont{Lepoutre}},
  \bibinfo{author}{\bibfnamefont{A.}~\bibnamefont{Aspect}}, \bibnamefont{and}
  \bibinfo{author}{\bibfnamefont{T.}~\bibnamefont{Bourdel}},
  \bibinfo{journal}{Phys. Rev. A} \textbf{\bibinfo{volume}{90}},
  \bibinfo{pages}{033405} (\bibinfo{year}{2014}),
  \urlprefix\url{https://link.aps.org/doi/10.1103/PhysRevA.90.033405}.

\bibitem[{Amp()}]{Amplitude}
\bibinfo{note}{It is noted however that for a {G}aussian fit, amplitude
  uncertainties are more strongly correlated to uncertainties in the width
  rather than uncertainties in resonance location.}

\bibitem[{\citenamefont{Beaufils et~al.}(2009)\citenamefont{Beaufils,
  Crubellier, Zanon, Laburthe-Tolra, Mar{\'{e}}chal, Vernac, and
  Gorceix}}]{2body}
\bibinfo{author}{\bibfnamefont{Q.}~\bibnamefont{Beaufils}},
  \bibinfo{author}{\bibfnamefont{A.}~\bibnamefont{Crubellier}},
  \bibinfo{author}{\bibfnamefont{T.}~\bibnamefont{Zanon}},
  \bibinfo{author}{\bibfnamefont{B.}~\bibnamefont{Laburthe-Tolra}},
  \bibinfo{author}{\bibfnamefont{E.}~\bibnamefont{Mar{\'{e}}chal}},
  \bibinfo{author}{\bibfnamefont{L.}~\bibnamefont{Vernac}}, \bibnamefont{and}
  \bibinfo{author}{\bibfnamefont{O.}~\bibnamefont{Gorceix}},
  \bibinfo{journal}{Physical Review A} \textbf{\bibinfo{volume}{79}},
  \bibinfo{pages}{032706} (\bibinfo{year}{2009}),
  \urlprefix\url{https://link.aps.org/doi/10.1103/PhysRevA.79.032706}.

\bibitem[{\citenamefont{Tiesinga et~al.}(1996)\citenamefont{Tiesinga, Williams,
  Julienne, Jones, Lett, and Phillips}}]{NaScattering}
\bibinfo{author}{\bibfnamefont{E.}~\bibnamefont{Tiesinga}},
  \bibinfo{author}{\bibfnamefont{C.~J.} \bibnamefont{Williams}},
  \bibinfo{author}{\bibfnamefont{P.~S.} \bibnamefont{Julienne}},
  \bibinfo{author}{\bibfnamefont{K.~M.} \bibnamefont{Jones}},
  \bibinfo{author}{\bibfnamefont{P.~D.} \bibnamefont{Lett}}, \bibnamefont{and}
  \bibinfo{author}{\bibfnamefont{W.~D.} \bibnamefont{Phillips}},
  \bibinfo{journal}{Journal of Research of the National Institute of Standards
  and Technology}  (\bibinfo{year}{1996}).

\bibitem[{sim()}]{simoniprivate}
\bibinfo{note}{A. Simoni, private communication}.

\bibitem[{\citenamefont{Ho and Shenoy}(1996)}]{Miscible1}
\bibinfo{author}{\bibfnamefont{T.-L.} \bibnamefont{Ho}} \bibnamefont{and}
  \bibinfo{author}{\bibfnamefont{V.~B.} \bibnamefont{Shenoy}},
  \bibinfo{journal}{Phys. Rev. Lett.} \textbf{\bibinfo{volume}{77}},
  \bibinfo{pages}{3276} (\bibinfo{year}{1996}),
  \urlprefix\url{https://link.aps.org/doi/10.1103/PhysRevLett.77.3276}.

\bibitem[{\citenamefont{Law et~al.}(1997)\citenamefont{Law, Pu, Bigelow, and
  Eberly}}]{Miscible2}
\bibinfo{author}{\bibfnamefont{C.~K.} \bibnamefont{Law}},
  \bibinfo{author}{\bibfnamefont{H.}~\bibnamefont{Pu}},
  \bibinfo{author}{\bibfnamefont{N.~P.} \bibnamefont{Bigelow}},
  \bibnamefont{and} \bibinfo{author}{\bibfnamefont{J.~H.}
  \bibnamefont{Eberly}}, \bibinfo{journal}{Phys. Rev. Lett.}
  \textbf{\bibinfo{volume}{79}}, \bibinfo{pages}{3105} (\bibinfo{year}{1997}),
  \urlprefix\url{https://link.aps.org/doi/10.1103/PhysRevLett.79.3105}.

\bibitem[{\citenamefont{Pu and Bigelow}(1998)}]{Miscible3}
\bibinfo{author}{\bibfnamefont{H.}~\bibnamefont{Pu}} \bibnamefont{and}
  \bibinfo{author}{\bibfnamefont{N.~P.} \bibnamefont{Bigelow}},
  \bibinfo{journal}{Phys. Rev. Lett.} \textbf{\bibinfo{volume}{80}},
  \bibinfo{pages}{1130} (\bibinfo{year}{1998}),
  \urlprefix\url{https://link.aps.org/doi/10.1103/PhysRevLett.80.1130}.

\bibitem[{\citenamefont{Lee et~al.}(2016)\citenamefont{Lee, J\o{}rgensen, Liu,
  Wacker, Arlt, and Proukakis}}]{ArltMisc}
\bibinfo{author}{\bibfnamefont{K.~L.} \bibnamefont{Lee}},
  \bibinfo{author}{\bibfnamefont{N.~B.} \bibnamefont{J\o{}rgensen}},
  \bibinfo{author}{\bibfnamefont{I.-K.} \bibnamefont{Liu}},
  \bibinfo{author}{\bibfnamefont{L.}~\bibnamefont{Wacker}},
  \bibinfo{author}{\bibfnamefont{J.~J.} \bibnamefont{Arlt}}, \bibnamefont{and}
  \bibinfo{author}{\bibfnamefont{N.~P.} \bibnamefont{Proukakis}},
  \bibinfo{journal}{Phys. Rev. A} \textbf{\bibinfo{volume}{94}},
  \bibinfo{pages}{013602} (\bibinfo{year}{2016}),
  \urlprefix\url{https://link.aps.org/doi/10.1103/PhysRevA.94.013602}.

\bibitem[{\citenamefont{Petrov}(2015)}]{Petrov}
\bibinfo{author}{\bibfnamefont{D.~S.} \bibnamefont{Petrov}},
  \bibinfo{journal}{Phys. Rev. Lett.} \textbf{\bibinfo{volume}{115}},
  \bibinfo{pages}{155302} (\bibinfo{year}{2015}),
  \urlprefix\url{https://link.aps.org/doi/10.1103/PhysRevLett.115.155302}.

\bibitem[{\citenamefont{Delannoy et~al.}(2001)\citenamefont{Delannoy, Murdoch,
  Boyer, Josse, Bouyer, and Aspect}}]{AspectSympathetic}
\bibinfo{author}{\bibfnamefont{G.}~\bibnamefont{Delannoy}},
  \bibinfo{author}{\bibfnamefont{S.~G.} \bibnamefont{Murdoch}},
  \bibinfo{author}{\bibfnamefont{V.}~\bibnamefont{Boyer}},
  \bibinfo{author}{\bibfnamefont{V.}~\bibnamefont{Josse}},
  \bibinfo{author}{\bibfnamefont{P.}~\bibnamefont{Bouyer}}, \bibnamefont{and}
  \bibinfo{author}{\bibfnamefont{A.}~\bibnamefont{Aspect}},
  \bibinfo{journal}{Phys. Rev. A} \textbf{\bibinfo{volume}{63}},
  \bibinfo{pages}{051602} (\bibinfo{year}{2001}),
  \urlprefix\url{https://link.aps.org/doi/10.1103/PhysRevA.63.051602}.

\bibitem[{\citenamefont{Ferrier-Barbut
  et~al.}(2016)\citenamefont{Ferrier-Barbut, Kadau, Schmitt, Wenzel, and
  Pfau}}]{Drop10}
\bibinfo{author}{\bibfnamefont{I.}~\bibnamefont{Ferrier-Barbut}},
  \bibinfo{author}{\bibfnamefont{H.}~\bibnamefont{Kadau}},
  \bibinfo{author}{\bibfnamefont{M.}~\bibnamefont{Schmitt}},
  \bibinfo{author}{\bibfnamefont{M.}~\bibnamefont{Wenzel}}, \bibnamefont{and}
  \bibinfo{author}{\bibfnamefont{T.}~\bibnamefont{Pfau}},
  \bibinfo{journal}{Phys. Rev. Lett.} \textbf{\bibinfo{volume}{116}},
  \bibinfo{pages}{215301} (\bibinfo{year}{2016}),
  \urlprefix\url{https://link.aps.org/doi/10.1103/PhysRevLett.116.215301}.

\bibitem[{\citenamefont{Schmitt et~al.}(2016)\citenamefont{Schmitt, Wenzel,
  B{ö}ttcher, Ferrier-Barbut, and Pfau}}]{Drop1}
\bibinfo{author}{\bibfnamefont{M.}~\bibnamefont{Schmitt}},
  \bibinfo{author}{\bibfnamefont{M.}~\bibnamefont{Wenzel}},
  \bibinfo{author}{\bibfnamefont{F.}~\bibnamefont{B{ö}ttcher}},
  \bibinfo{author}{\bibfnamefont{I.}~\bibnamefont{Ferrier-Barbut}},
  \bibnamefont{and} \bibinfo{author}{\bibfnamefont{T.}~\bibnamefont{Pfau}},
  \bibinfo{journal}{Nature 539, 259}  (\bibinfo{year}{2016}).

\bibitem[{\citenamefont{Chomaz et~al.}(2016)\citenamefont{Chomaz, Baier,
  Petter, Mark, W\"achtler, Santos, and Ferlaino}}]{Drop2}
\bibinfo{author}{\bibfnamefont{L.}~\bibnamefont{Chomaz}},
  \bibinfo{author}{\bibfnamefont{S.}~\bibnamefont{Baier}},
  \bibinfo{author}{\bibfnamefont{D.}~\bibnamefont{Petter}},
  \bibinfo{author}{\bibfnamefont{M.~J.} \bibnamefont{Mark}},
  \bibinfo{author}{\bibfnamefont{F.}~\bibnamefont{W\"achtler}},
  \bibinfo{author}{\bibfnamefont{L.}~\bibnamefont{Santos}}, \bibnamefont{and}
  \bibinfo{author}{\bibfnamefont{F.}~\bibnamefont{Ferlaino}},
  \bibinfo{journal}{Phys. Rev. X} \textbf{\bibinfo{volume}{6}},
  \bibinfo{pages}{041039} (\bibinfo{year}{2016}),
  \urlprefix\url{https://link.aps.org/doi/10.1103/PhysRevX.6.041039}.

\bibitem[{\citenamefont{Cabrera et~al.}(2017)\citenamefont{Cabrera, Tanzi,
  Sanz, Naylor, Thomas, Cheiney, and Tarruell}}]{Drop3}
\bibinfo{author}{\bibfnamefont{C.~R.} \bibnamefont{Cabrera}},
  \bibinfo{author}{\bibfnamefont{L.}~\bibnamefont{Tanzi}},
  \bibinfo{author}{\bibfnamefont{J.}~\bibnamefont{Sanz}},
  \bibinfo{author}{\bibfnamefont{B.}~\bibnamefont{Naylor}},
  \bibinfo{author}{\bibfnamefont{P.}~\bibnamefont{Thomas}},
  \bibinfo{author}{\bibfnamefont{P.}~\bibnamefont{Cheiney}}, \bibnamefont{and}
  \bibinfo{author}{\bibfnamefont{L.}~\bibnamefont{Tarruell}},
  \bibinfo{journal}{arXiv:1708.07806}  (\bibinfo{year}{2017}).

\bibitem[{\citenamefont{Cheiney et~al.}(2017)\citenamefont{Cheiney, Cabrera,
  Sanz, Naylor, Tanzi, and Tarruell}}]{Drop4}
\bibinfo{author}{\bibfnamefont{P.}~\bibnamefont{Cheiney}},
  \bibinfo{author}{\bibfnamefont{C.~R.} \bibnamefont{Cabrera}},
  \bibinfo{author}{\bibfnamefont{J.}~\bibnamefont{Sanz}},
  \bibinfo{author}{\bibfnamefont{B.}~\bibnamefont{Naylor}},
  \bibinfo{author}{\bibfnamefont{L.}~\bibnamefont{Tanzi}}, \bibnamefont{and}
  \bibinfo{author}{\bibfnamefont{L.}~\bibnamefont{Tarruell}}
  (\bibinfo{year}{2017}), \eprint{1710.11079v1}.

\bibitem[{\citenamefont{Semeghini et~al.}(2017)\citenamefont{Semeghini,
  Ferioli, Masi, Mazzinghi, Wolswijk, Minardi, Modugno, Modugno, Inguscio, and
  Fattori}}]{Drop5}
\bibinfo{author}{\bibfnamefont{G.}~\bibnamefont{Semeghini}},
  \bibinfo{author}{\bibfnamefont{G.}~\bibnamefont{Ferioli}},
  \bibinfo{author}{\bibfnamefont{L.}~\bibnamefont{Masi}},
  \bibinfo{author}{\bibfnamefont{C.}~\bibnamefont{Mazzinghi}},
  \bibinfo{author}{\bibfnamefont{L.}~\bibnamefont{Wolswijk}},
  \bibinfo{author}{\bibfnamefont{F.}~\bibnamefont{Minardi}},
  \bibinfo{author}{\bibfnamefont{M.}~\bibnamefont{Modugno}},
  \bibinfo{author}{\bibfnamefont{G.}~\bibnamefont{Modugno}},
  \bibinfo{author}{\bibfnamefont{M.}~\bibnamefont{Inguscio}}, \bibnamefont{and}
  \bibinfo{author}{\bibfnamefont{M.}~\bibnamefont{Fattori}}
  (\bibinfo{year}{2017}), \eprint{1710.10890v1}.

\end{thebibliography}

\end{document}